\documentclass[12pt,letterpaper]{article}
\usepackage{graphicx,array}
\usepackage{color}
\usepackage{latexsym}
\usepackage{amsthm}
\usepackage{amsmath}
\usepackage{amssymb}
\usepackage[numbers,sort&compress]{natbib}
\usepackage{bm}
\usepackage{slashed}
\usepackage{tensor}
\usepackage{mathrsfs}
\usepackage{url}
\usepackage{hyperref} %Automatically links \label and \ref commands; Always load last
\hypersetup{
    colorlinks=true,       % false: boxed links; true: colored links
    linkcolor=red,          % color of internal links
    citecolor=blue,        % color of links to bibliography
    filecolor=magenta,      % color of file links
    urlcolor=blue           % color of external links
}
\usepackage[all]{hypcap} %Link navagates to top of figure instead of caption (below fig)

\usepackage[inline]{enumitem}
\usepackage{multirow}
\usepackage{pdflscape}

\usepackage{myterms}
\usepackage{latexsym}

\newcommand*{\vv}[1]{\vec{\mkern0mu#1}}
\newcommand{\GN}{G_{\! N}}
\newcommand{\nc}{\newcommand}
\nc{\beq}{\begin{equation}}
\nc{\eeq}{\end{equation}}
\nc{\beqa}{\begin{eqnarray}}
\nc{\eeqa}{\end{eqnarray}}
\nc{\bit}{\begin{itemize}}
\nc{\eit}{\end{itemize}}
\nc{\mc}{\mathcal}
\newcommand{\RE}{R_\mathrm{E}}
\newcommand{\RS}{R_\mathrm{S}}
\newcommand{\RB}{R_\mathrm{B}}
\newcommand{\Rrel}{r_\mathrm{rel}}
\newcommand{\Rph}{r_\mathrm{ph}}
\newcommand{\Rstart}{r_\mathrm{start}}
\newcommand{\Rend}{r_\mathrm{end}}
\newcommand{\tB}{t_\mathrm{B}}
\newcommand{\tH}{t_\mathrm{H}}
\newcommand{\PBH}{\text{\sc pbh}}
\newcommand{\dash}{\text{--}}

\numberwithin{equation}{section}

\title{\bf Tests of Dark MACHOs:  \\ Lensing, Accretion, and Glow
}

\author{\large Yang Bai$^{a}$, Andrew J. Long$^{b}$, and Sida Lu$^{c}$}
\date{\small \it 
$^a$Department of Physics, University of Wisconsin-Madison, Madison, WI 53706, USA \\
$^b$Department of Physics and Astronomy, Rice University, Houston, TX 77005, USA \\
$^c$Department of Physics, Tel Aviv University, Tel-Aviv 69978, Israel 
}

\begin{document}

\maketitle

\setlength{\parskip}{0.2ex}

\begin{abstract}
Dark matter could take the form of dark massive compact halo objects (dMACHOs); \textit{i.e.}, composite objects that are made up of dark-sector elementary particles, that could have a macroscopic mass from the Planck scale to above the solar mass scale, and that also admit a wide range of energy densities and sizes. Concentrating on the gravitational interaction of dMACHOs with visible matter, we map out the mass-radius parameter space that is consistent with gravitational lensing experiments, as well as anisotropies of the cosmic microwave background (CMB) based on the spherical accretion of matter onto a dMACHO in the hydrostatic approximation. For dMACHOs with a uniform-density mass profile and total mass in the range of $\sim 10^{-12} - 10\,M_\odot$, we find that a dMACHO could explain 100\% of the dark matter if its radius is above $\approx 3$ times the Einstein radius of the lensing system. For a larger mass above $10\,M_\odot$, a dMACHO with radius above $\sim 1 \times 10^8 \cm \times(M/100\,M_\odot)^{9/2}$ is consistent with CMB observables.  For a lighter dMACHO with mass below $\sim 10^{-12}\,M_\odot$, there still is not a good experimental probe. Finally, we point out that heavier dMACHOs with masses $\sim 0.1\,M_\odot$ may be observed by X-ray and optical telescopes if they reside at rest in a large molecular cloud, nearby to our solar system, and accrete ordinary matter to emit photons.
 \end{abstract}

\newpage

\begingroup
\hypersetup{linkcolor=black,linktocpage}
\tableofcontents 
\endgroup
\newpage

%==================================
% Introduction
%==================================
\section{Introduction}\label{sec:introduction}

%=========
While the presence of dark matter in our Universe is firmly established by an abundance of empirical evidence, its properties and interactions remain a mystery.  
Many studies have explored the idea that dark matter may be a collection of point-like (elementary) particles~\cite{Bertone:2004pz}.  
To explain the dark matter's apparent gravitational interactions with visible matter, viable candidates for particle dark matter should have nonzero mass.  
Moreover, elementary particle masses are generally bounded from above by the Planck mass scale, $\Mpl \simeq 2.43 \times 10^{18} \GeV / c^2 \simeq 4.34 \times 10^{-6} \gram$, since a heavier particle would collapse to form a black hole and evaporate quickly.  
On the other hand, it is not hard to have candidates for dark matter with masses exceeding the Planck mass scale: $M > M_\mathrm{pl}$. Such dark matter candidates certainly cannot be elementary particles, but rather they must be composite objects with a size $R \gg \hbar c/M$, much larger than the Compton wavelength.  

%=========
Historically, massive astrophysical  compact halo objects (MACHOs) were proposed as one of the earliest solutions of the dark matter problem~\cite{Paczynski:1985jf,Griest:1990vu} (see Ref.~\cite{Bertone:2016nfn} for a recent review).  ``MACHO'' can refer to any macroscopic object composed of standard matter, but which does not glow and thereby evades detection from all but gravitational probes.  MACHO candidates include planets and ``dead stars'' such as brown or red dwarfs~\cite{Griest:1990vu}; masses can range anywhere from $\sim 10^{-6}$ to $10$ solar masses while the corresponding sizes vary in accordance with a typical atomic energy density $\rho \sim 1 \gram/\mathrm{cm}^3$.  
However, MACHOs composed of ordinary (baryonic) matter were ultimately excluded as viable dark matter candidates because observations of the cosmic microwave background and the abundances of light elements provided measurements of the cosmological baryon density, which was remarkably consistent with the observed amount of luminous baryonic matter.~\footnote{One caveat is that MACHOs could have a primordial formation history before Big Bang nucleosynthesis (BBN), \textit{e.g.} the quark nugget~\cite{Witten:1984rs}. }

%=========
Despite the shortcomings of MACHO dark matter, a variety of compelling theories predict a new class of macroscopic dark matter states that are made of particles in the dark sector, or a sector different from our Standard Model (SM) sector. We will use ``dark-MACHOs'' or ``dMACHOs'' as a phenomenological catch-all term to denote any macroscopic dark matter candidate ($R \gg \hbar c / M$) that interacts predominantly gravitationally with standard matter.  
Examples of dMACHOs include primordial black holes, $Q$-balls and other solitons, quark nuggets, asymmetric dark matter nuggets~\cite{Wise:2014jva,Gresham:2017cvl}, dark blobs~\cite{Grabowska:2018lnd}, N-MACHOs~\cite{Dvali:2019ewm}, and mirror stars~\cite{Curtin:2019ngc}.  
Despite their unusually large mass and size, the models underlying dMACHOs are generally no more baroque than many models of particle dark matter, and in fact the cosmological production of dMACHOs can typically be accomplished with only minimal interactions.  

\begin{figure}[thb!] %[tbp]
\centering
\includegraphics[width=0.65\linewidth]{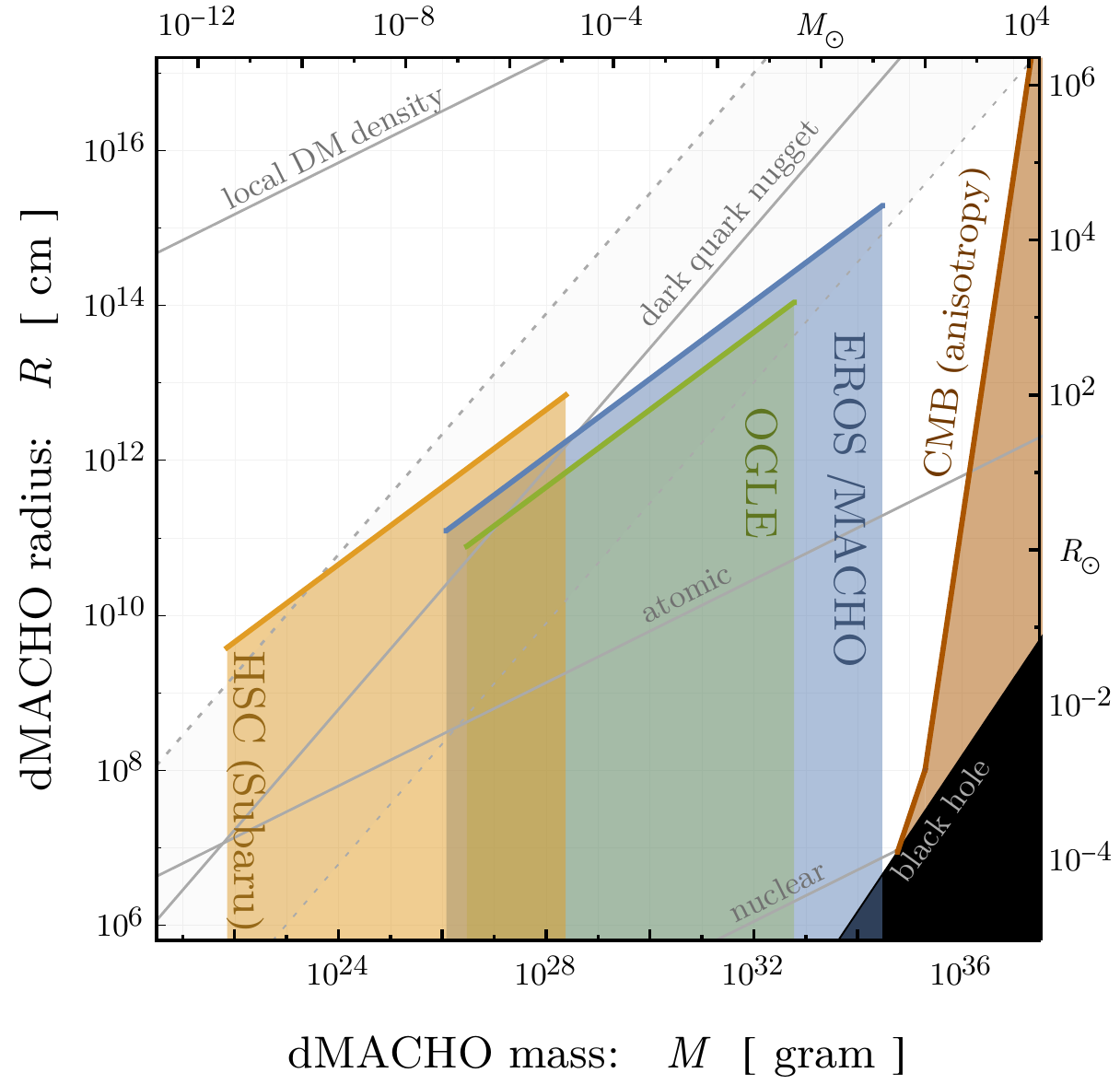} 
\caption{The phenomenological dMACHO parameter space is shown in terms of the dMACHO's mass $M$ and scale radius $R$. Here, dMACHOs are assumed to compose 100\% of dark matter. Tests of dMACHOs (gravitational lensing and CMB anisotropy from baryon accretion) lead to an exclusion of the shaded regions. For comparison, we also show the black hole Schwarzschild radius, lines of constant energy density -- nuclear ($10 \MeV / \mathrm{fm}^3$), atomic ($1 \gram /\mathrm{cm}^3$), and local dark matter energy density ($0.4 \GeV/\mathrm{cm}^3$) --  and the expected mass-radius relation for dark quark nuggets~\cite{Bai:2018dxf}.  
%\SL{The window between $1 \dash 100 \Msun$ could possibly be covered by other searching methods like the supernovae probability density function~\cite{Bernal:2017vvn,Zumalacarregui:2017qqd}.
%\SL{The window between $1 \dash 100 \Msun$ is also probed by other systems, not considered here, including dwarf galaxies~\cite{Carr:1997cn} and supernovae lensing~\cite{Bernal:2017vvn,Zumalacarregui:2017qqd}.}
}\label{fig:mass_radius}
\end{figure}

%=========
Macroscopic dark matter candidates have drawn increasing attention recently. 
Various detection methods have been introduced, including lensing, X-ray emission and thermonuclear interaction~\cite{Inoue:2017csr,Sidhu:2019kpd,Croon:2020wpr}.  
In this article we study various probes of dMACHOs and will treat its mass $M$ and radius $R$ as two phenomenological parameters.  
For simplicity, we only study spherically symmetric dMACHOs here.  
In particular we investigate the following. 
\begin{itemize}
	\item  Gravitational lensing.  The presence of dMACHOs in the Milky Way halo can induce a gravitational lensing of background stars.  Telescopes such as Subaru/HSC, EROS/MACHO, and OGLE have searched for evidence of this lensing, and by not finding any convincing effect, they constrain the MACHO parameter space.  We recast these limits into the phenomenological dMACHO parameter space, described by the characteristic mass $M$ and radius $R$.  
	\item  Accretion.  In baryon-dense environments dMACHOs may accrete ordinary matter though the force of gravity.  As the accreted matter is heated, the dMACHO develops a glowing halo.  
The accretion of baryonic matter onto dMACHOs in the early universe is shown to change the ionization fraction and the cosmic microwave background (CMB) anisotropy, similar to the analogous effect with primordial black holes (PBHs).  We also study the prospects for identifying such dMACHO halos in the Milky Way today, and we find that for some range of dMACHO masses, telescopes could identify a dMACHO as it travels with a small speed inside a molecular cloud. 
\end{itemize} 
The constraints of PBHs as dark matter candidates from the above two aspects has been studied in several references~\cite{Carr:2016drx,Ali-Haimoud:2016mbv,Serpico:2020ehh}, and there is also similar discussion for extended astrophysical objects~\cite{Savastano:2019zpr}.  
The difference of our analysis from the previous articles is the accretion profile derived in our analysis. 
Specifically, we consider in this work the accretion profiles based on the hydrostatic approximation, while the Bondi accretion profiles are used in most of the previous articles. We will discuss more about why we choose the hydrostatic approximation for dMACHOs and the difference between the Bondi and the hydrostatic accretions.
We summarize our main results based on these tests in \fref{fig:mass_radius} in terms of the dMACHO's mass $M$ and characteristic radius $R$.

%=========
The remainder of this article is organized as follows.  
In Sec.~\ref{sec:candidates} we list several dMACHO candidates and their properties.
In Secs.~\ref{sec:lensing}~and~\ref{sec:accretion} we discuss how gravitational lensing and accretion can be used to test dMACHOs.  
We summarize our results and conclude in \sref{sec:conclusion}.  
For the remainder of the article we take $\hbar = c = k_B = 1$.  
In these units the reduced Planck mass is $\Mpl \simeq 2.43 \times 10^{18} \GeV$ and Newton's constant is $\GN = 1 / (8 \pi \Mpl^2) \simeq 6.71 \times 10^{-39} \GeV^{-2}$.  

%==================================
% Dark MACHO candidates
%==================================
\section{Dark MACHO candidates}\label{sec:candidates}

%=========
In this section we list several candidates for dMACHOs.  
The formation process and mechanism of these candidates (see \textit{e.g.} Ref.~\cite{Chang:2018bgx}) are not directly related to our phenomenological analysis later.  
Each candidate is characterized by a mass $M$, a scale radius $R$, and an enclosed mass profile $\widetilde{M}(r)$, such that $\widetilde{M}(r)$ is the mass contained within a sphere of radius $r = |\vec{r}|$ centered at the dMACHO, and $\widetilde{M}$ rapidly approaches $M$ for $r > R$.  

%=========
\paragraph*{Primordial black hole.}  
Perhaps the most well studied example of a dMACHO is PBH dark matter~\cite{Carr:1974nx,Carr:2016drx}.  
A PBH corresponds to a dMACHO whose mass is contained within its Schwartzchild radius, $\RS = 2 \, \GN M$.  
A PBH can be treated as a point mass on larger length scales, and its mass density can be written as $\rho({\bm x}) = M \, \delta^3({\bm x})$. The enclosed mass function has $\widetilde{M}(r) = M$ for $r > \RS$. While our results will apply also to the case of a PBH, we are primarily interested in extended dMACHOs for which $R \gg \RS$.  

%=========
\paragraph*{Quark nuggets.}  
Another class of compelling dMACHO candidates includes the QCD quark nugget~\cite{Witten:1984rs}, the axion quark nugget~\cite{Zhitnitsky:2002qa}, the six-flavor quark nugget~\cite{Bai:2018vik}, and the dark quark nugget~\cite{Bai:2018dxf}.  
These compact objects contain a gas of interacting fermions in the unconfining phase supported by their degeneracy pressure to balance the external vacuum pressure. 
The energy density on the interior of the nugget is approximately uniform and its magnitude is set by the confinement scale $\Lambda$ of the strong interaction: $\rho \sim M/R^3 \sim \Lambda^4$. For spherically-symmetric dMACHOs with a uniform density on their interior, we can write the mass profile as 
\begin{align}\label{eq:hard_sphere}
	\widetilde{M}(r) = 
	\begin{cases}
	M \left( \dfrac{r}{R} \right)^3 & , \ r \leq R ~, \\
	M & , \ r > R  ~.
	\end{cases}
\end{align}
We will often use \eref{eq:hard_sphere} for our phenomenological studies in the following sections.

%=========
\paragraph*{$\bm{Q}$-balls and non-topological solitons.} 

Similar to quark nuggets that are made of fermions, another class of dMACHO candidates includes $Q$-balls~\cite{Coleman:1985ki} and non-topological solitons~\cite{Lee:1991ax} that are made of a complex scalar boson with an unbroken $\mathrm{U}(1)$ symmetry to ensure its stability. Either the scalar quantum pressure or its self-interaction is responsible for balancing the external vacuum pressure (or gravitational pressure for a very heavy one). If the self-interaction is not important, the scalar field profile has a core structure with the enclosed mass profile as~\cite{Ponton:2019hux}
\begin{align}
	\widetilde{M}(r) \approx 
	\begin{cases}
	M \left[ \dfrac{r}{R} - \dfrac{\sin{(\frac{2\pi r}{R})}}{2\pi} \right] & , \ r \leq R ~, \\
	M & , \ r > R  ~.
	\end{cases}
\end{align}
In the innermost region with $r \ll R$, the mass profile has the same scaling as in \eref{eq:hard_sphere}, namely $\widetilde{M} \propto r^3$.  If the self-interaction is important, the scalar field profile has a step-function behavior, and the enclosed mass profile is given by \eref{eq:hard_sphere}.  

The axion star is another well-studied dark matter candidate that falls into this category of dMACHO models. For a diluted axion star (the axion self-interaction is not important), the quantum pressure is balanced by the gravitational pressure~\cite{Kolb:1993zz}. The energy density profile follows the exponential 1$S$ state of hydrogen, $\propto e^{-r/R}$, and the enclosed mass is
\begin{align}
	\widetilde{M}(r) = 
	M \left[ 1-  \left( 1 + \dfrac{r}{R} + \dfrac{r^2}{2R^2} \right) e^{- r/R}\right]& ~.
\end{align}

%=========
\paragraph*{Ultra-compact mini-halo.}  
For relatively small primordial density perturbations $3\times 10^{-4}<\delta<0.3$, gravity might be too weak to enable the formation of a primordial black hole. Nevertheless, it is possible that such a spatial over-density can still seed the growth of some ultra-compact mini-halos~\cite{Ricotti:2009bs}. The density profile of a mini-halo is predicted to have a power-law behavior of $\rho(r)\propto r^{-9/4}$ from secondary infall~\cite{Bertschinger:1985pd}, where the large power-law index renders a significantly larger density at the core. Their enclosed mass profile is thus
\begin{align}\label{eq:UCMH}
\widetilde{M}(r) = 
	\begin{cases}
	M \left( \dfrac{r}{R} \right)^{3/4} & , \ r \leq R ~, \\
	M & , \ r > R  ~,
	\end{cases}
\end{align}
where $R$ is the radius of the mini-halo.

%=========
The mass profiles listed above are the properties for individual dMACHOs, and to derive constraints on these models one will also need to know the distribution of dMACHO masses and radii across the population.  We assume that all dMACHOs are identical with the same mass profile, rather than introducing a distribution over masses and radii.  This assumption is primarily a matter of simplicity, and it would be interesting to explore more general mass and radius distributions, as has been done for studies of PBH dark matter~\cite{Bernal:2017vvn,Kuhnel:2017pwq,Carr:2017jsz}.  
%Following the approach that is often employed for studies of PBHs, we assume that all dMACHOs are identical with the same mass profile.  
%In practice it is usually assumed that the dMACHOs are monochromatic in mass, which is also the approach we will take in the following analysis.

%==================================
% Lensing
%==================================
\section{Gravitational lensing}\label{sec:lensing}

%=========
Gravitational lensing provides an effective probe of mass distributions throughout the universe.  
For most applications, it is customary to assume that the lensing mass distribution is point-like and then the effective ``size'' of the lens is given by the Einstein radius, $\RE$.  
However, since the dMACHO is an extended object with scale radius $R$, the point-like lens approximation breaks down for $R \gtrsim \RE$, and in general the magnification is reduced with respect to the point-like lens calculation.  
In this section we recast constraints from existing lensing experiments onto dMACHOs.

%==================================
% Lensing by a point mass lens
%==================================
\subsection{Lensing by a point mass lens}\label{sub:point_mass_lens}

%=========
We begin by reviewing the well-known results for gravitational lensing by a point-like mass distribution.  
The quantity of interest throughout this section is the magnification factor $\mu$ that quantifies how much a light source is brightened ($\mu > 1$) or diminished ($0 < \mu < 1$) by an intervening lens.  
In the geometrical optics approximation, the magnification factor that results from a point-like mass distribution is given by a simple formula:~\cite{Griest:1990vu}
\beqa
\label{eq:mu-point-mass}
\mu = \frac{y^2 + 2}{y\,(y^2 + 4)^{1/2} } \com
\eeqa
where $y \equiv d_\mathrm{S}/\RE$ is the dimensionless source position on the lens plane, $d_\mathrm{S}$ is the tangential distance between the source and lens in the lens plane, and $\RE$ is the Einstein radius.  
The Einstein radius is  
\beqa
\label{eq:Einstein-radius}
\RE  = \sqrt{4\, \GN \, M\, \kappa(1-\kappa)\, D_\mathrm{OS} } \approx (1.51 \times 10^{14} \ \mathrm{cm}) \times \left( \frac{\sqrt{\kappa(1-\kappa)}}{1/2} \right) \, \left( \frac{D_\mathrm{OS}}{50 \kpc}\right)^{1/2} \,  \left( \frac{M}{M_\odot}\right)^{1/2}  \,,
\eeqa
where $\GN$ is Newton's constant, $M$ is the lensing mass, $\kappa \equiv D_\mathrm{OL}/D_\mathrm{OS}$ is a dimensionless ratio, $D_\mathrm{OL}$ is the angular diameter distance between the observer and the lens, and $D_\mathrm{OS}$ is the distance between the observer and the source.  
The solar mass in grams is $M_\odot = 1.989 \times 10^{33} \gram$; the Earth-Sun distance in meters is $1 \, \mathrm{AU} =1.496 \times 10^{13} \ \mathrm{cm}$; and a kiloparsec in meters is $1 \kpc = 3.086 \times 10^{21} \ \mathrm{cm}$.  
We will see below how \eref{eq:mu-point-mass} is generalized for an extended lensing mass.  

%=========
Throughout this section we will use the geometrical optics approximation.  
If this approximation is reliable for photons of energy $E_\gamma$ then one requires $4 \GN M E_\gamma \gtrsim 1$~\cite{Gould1992ApJ}.  
At the end of this section we derive constraints on $M$ from optical lensing surveys for which $E_\gamma \gtrsim 1 \eV$, and the geometrical optics approximation will be applicable\footnote{For smaller $M$ the wave-like nature of light must be taken into account when calculating the magnification.  For instance, a recent study by the Subaru/HSC telescope~\cite{Niikura:2017zjd}, found that their limit on PBH-induced lensing weakened quickly outside of the geometrical optics regime for $M \lesssim 10^{22} \gram$.} as long as $M \gtrsim (6.6 \times 10^{22} \gram) \times (1 \eV/E_\gamma)$.
%If this approximation is reliable for photons of energy $E_\gamma$ then one requires $4 \GN M E_\gamma \gtrsim 1$~\cite{Gould1992ApJ}; \textit{i.e.}, a heavier $M$ or higher photon energy is needed.  
%Ignoring the redshift effects, validity of the optics approximation sets a lower bound on the dMACHO's mass, $M \gtrsim (6.6 \times 10^{22} \gram) \times (1 \eV/E_\gamma)$, for photons in the visible spectrum.  
%\SL{Since the experiments constraints we will recast from are mostly on the optical band\footnote{The Subaru/HSC result takes the wave effects into account~\cite{Niikura:2017zjd}, but the constraint becomes quickly weaken after entering the mass regime of wave optics.} and are in the mass regime of above around $10^{22} \gram$, we will limit ourself to focus on $E_\gamma\gtrsim 1 \eV$. For lower energy there also corresponding searches, e.g. in the radio band using the pulsar timing~\cite{Dror:2019twh}.}
%In this paper, we will concentrate on the region with the dMACHO mass above around $10^{22} \gram$.  

%==================================
% Generalization to axially-symmetric lenses
%==================================
\subsection{Generalization to axially-symmetric lenses}\label{sub:lens_formulas}

\begin{figure}[thb!] %[tbp]
\centering
%\vspace{-10mm}
\includegraphics[width=0.75\linewidth]{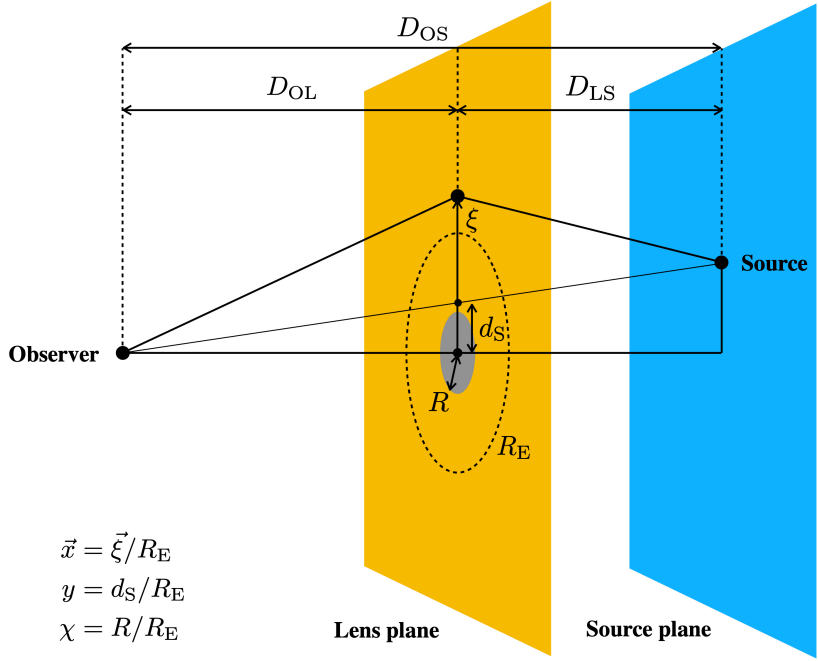} 
%\vspace{-5mm}
\caption{This diagram shows the geometry of the dMACHO lensing system and defines several dimensionless variables that are used in the text, namely $\vec{x}$, $y$, and $\chi$.  
}\label{fig:lensing_notation}
\end{figure}

%=========
In this section we generalize the previous results for point-like lenses to accommodate any axially-symmetric lenses.  
In particular, we are interested in dMACHOs with spherical mass distributions that corresponds to an axially-symmetric lens after projecting the mass distribution into the lens plane. 
For geometrical lensing, we follow the lecture notes in Refs.~\cite{Narayan:1996ba,Meneghetti:book}.  
The geometrical setup and some important notation are shown in \fref{fig:lensing_notation}.

%=========
For a general lens and in the thin screen approximation, the lens can be described by a planar distribution of matter. 
Let $\rho(\vv{\xi}, z)$ be the lens's mass density where $\vv{\xi}$ is a two-dimensional vector in the lens plane and $z$ is the coordinate normal to the plane.  
The lens's surface density is 
\beqa
\Sigma(\vv{\xi}) = \int \! \ud z \ \rho(\vv{\xi}, z) \per
\eeqa
The lens's mass induces a gravitational potential $\Phi(\vv{\xi}, z)$, which is the three-dimensional Newtonian potential satisfying the Poisson equation $\nabla^2 \Phi = 4 \pi \GN\, \rho$. 
Using the Einstein radius as a unit, one can define a dimensionless {\it effective lensing potential} as
\beqa
\Psi(\vv{\xi}) \equiv 2\,\frac{D_\mathrm{OL}\,D_\mathrm{LS}}{\RE^2\,D_\mathrm{OS}} \int \! \ud z \, \Phi(\vv{\xi}, z) \com
\eeqa
where $D_\mathrm{LS}$ is the angular diameter distance from the lens to the source.  
In terms of the dimensionless vector $\vv{x} = \vv{\xi}/R_\mathrm{E}$, the ``Jacobian matrix" of $\Psi(\vec{x})$ is
\beqa
\mathcal{J}_{ij} = \left( \delta_{ij} - \frac{\partial^2 \Psi(\vv{x})}{\partial x_i \partial x_j} \right) ~.
\eeqa
The magnification tensor or matrix is just its inverse, $\mathcal{M} = \mathcal{J}^{-1}$.  
The magnification factor for any individual image $i$ is simply $\mu_i = \mbox{det}(\mathcal{M})_i$, evaluated at the image's position on the lens plane. 

%=========
For an axially-symmetric lens with the optical axis along the lens's center, one has $\Sigma(\vv{\xi}) = \Sigma(|\vv{\xi}|)$.  
For such a system, the images appear along the line where the lens plane intersects the plane containing the source and the optical axis; we use the dimensionless variable $x = |\vv{\xi}|/\RE$ to measure distances along this line.  
The dimensionless deflection angle is
\beqa
\qquad \qquad \alpha(x) = \frac{m(x)}{x} ~, \qquad \qquad  \mbox{with}\, \qquad m(x) \equiv \frac{\RE^2}{M}\, \int^x_0 \! \ud x' \ 2\pi\,x' \, \Sigma(x') \per
\eeqa
In this notation the lens equation is simply
\beqa
\label{eq:lens}
y = x - \alpha(x) = x - \frac{m(x)}{x} ~,
\eeqa
where $y = d_\mathrm{S} / \RE$, which was defined below \eref{eq:mu-point-mass}, is the dimensionless source position on the lens plane.  
In general the lens equation may have multiple solutions, $x_i$, corresponding to different images in the lens plane. Furthermore, some solutions could have $x_i < 0$, which should be interpreted to mean that the corresponding image is on the opposite side of the lens center compared to the source. The magnification factor of the $i^\mathrm{th}$ image is given by 
\beqa
\label{eq:magnification-formula}
\mu_i = \mbox{det}(\mathcal{M})_i = \frac{1}{\mbox{det}(\mathcal{J})_i} = \left(\frac{y}{x}\frac{dy}{dx}\right)_i^{-1} =  \left(1 - \frac{\alpha(x)}{x} \right)_i^{-1}\, \left(1 - \frac{d\alpha(x)}{dx} \right)_i^{-1} ~,
\eeqa
and the total magnification factor is
\beqa
\mu \equiv \sum_i \ \lvert\mu_i \rvert  ~. 
\eeqa
For instance, if the lens can be approximated as a point mass, one finds $m(x)=1$ and $\alpha(x)=1/x$, leading to the magnification factor formula in \eref{eq:mu-point-mass}.  

%==================================
% Lensing magnification from dMACHOs
%==================================
\subsection{Lensing magnification from dMACHOs}\label{sub:mag}

\begin{figure}[p] %[tbp]
	\centering
		\includegraphics[width=0.47
	\linewidth]{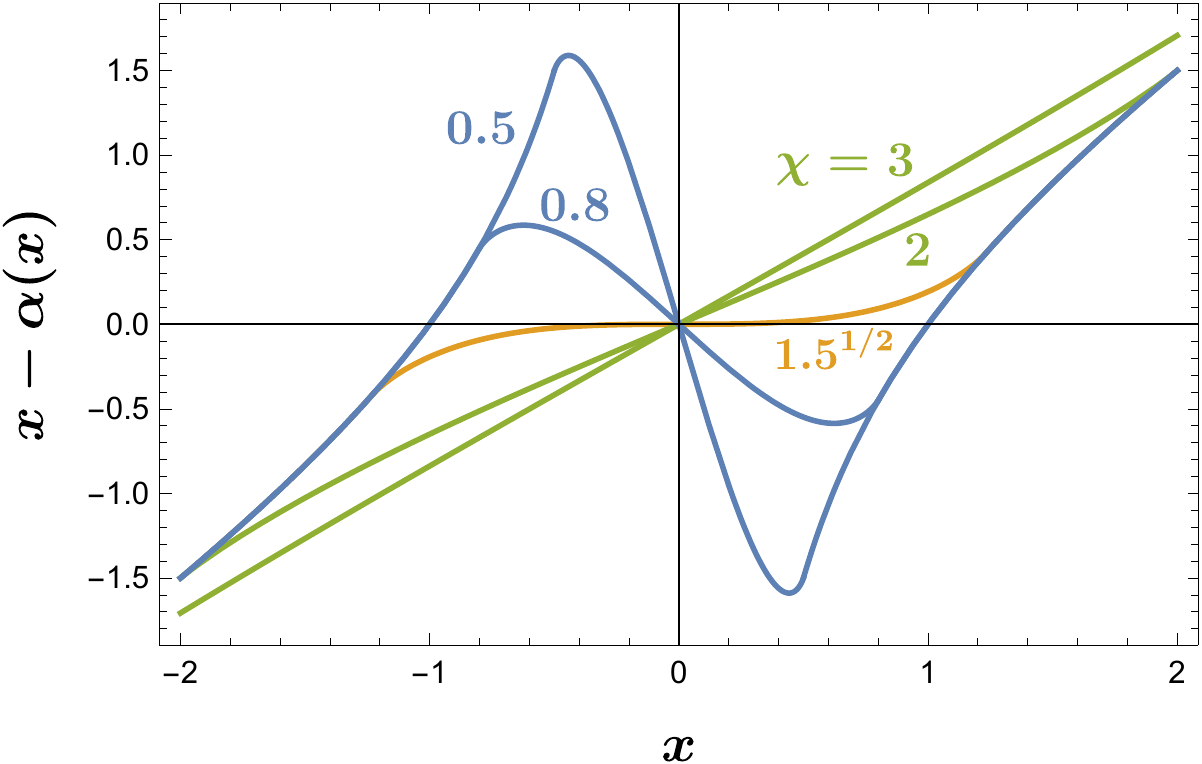} \quad \hspace{3mm}
	\includegraphics[width=0.45\linewidth]{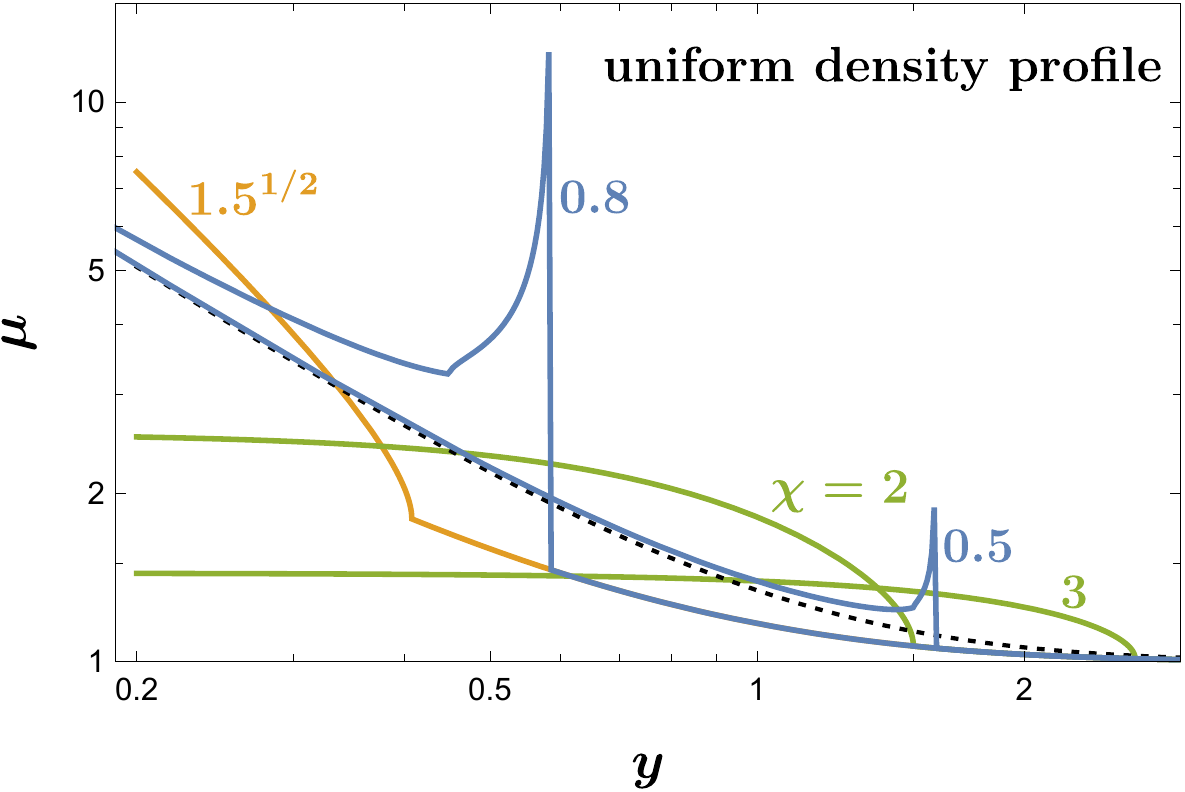} 
	\caption{\textit{Left panel:} We show the function $x- \alpha(x)$ for different values of $\chi = R/\RE$.  Solving $y = x - \alpha(x)$ determines the number and location of images.  We assume a uniform dMACHO mass density profile.  \textit{Right panel:} We show the magnification factor $\mu$ as a function of $y$ for several values of $\chi$. The black and dotted line is the function for a point-like mass in \eref{eq:mu-point-mass}. The spikes for $\chi=0.5$ and $0.8$ are due to caustic crossing with $d\alpha/dx=1$ in \eref{eq:magnification-formula}. 
	 }
	\label{fig:xalpha-hard}
\end{figure}

%=========
If the spatial extension of the lensing mass is relevant, then the formalism above can be used to derive an expression for the magnification factor.  
Let us first consider a uniform density dMACHO as in \eref{eq:hard_sphere} that has a constant mass density inside of a radius $R$.  
The function $m(x)$ is calculated to be 
\beqa
m(x)^{\rm uniform}  = 
\begin{cases}
  1 - \left(1 - \dfrac{x^2}{\chi^2}\right)^{3/2} ~, & \quad \text{for } x \leq \chi \\
    1   ~,         &   \quad \text{for } x > \chi  
\end{cases} \com
\eeqa
where $\chi \equiv R/\RE$ is a dimensionless measure of the dMACHO's size; we are generally interested in dMACHOs with $\chi < O(1)$.  
The left panel of \fref{fig:xalpha-hard} shows $x-m(x)/x$.  
For a large value of $\chi \geq \sqrt{3/2}$, the function $x - m(x)/x$ is monotonically increasing; in this case, the lens equation \pref{eq:lens} has only one solution for any $y$, and only one image is anticipated.  
For a smaller value of $\chi < \sqrt{3/2}$, there could be one image or three images depending on the value of $y$.  

%=========
The magnification factor $\mu$ is calculated using \eref{eq:magnification-formula}.  
In the right panel of \fref{fig:xalpha-hard}, we show the magnification factor as a function of the source position $y$ for different dMACHO radii, $\chi$.  
When the dMACHO radius decreases, $\chi \rightarrow 0$, the magnification approaches the well-known result for a point-like lens \pref{eq:mu-point-mass}.  
For a given $\chi < 1$, there a certain value of $y$ at which $\mu$ grows sharply.  
This value of $y$ corresponds to the turning point of $x - \alpha(x)$ at $x < 0$ in the left panel, and the large magnification factor can be seen from \eref{eq:magnification-formula} since the first derivative of $x-\alpha(x)$ vanishes at the turning point.  
For $\chi > \sqrt{3/2}$ the magnification is largest as $y \to 0$, where it can be approximated as 
\beqa\label{eq:mu_max_1}
\mu_\mathrm{max}^{\rm uniform} 
\approx  \left(1 - \dfrac{3}{2\,\chi^2}\right)^{-2} \per
\eeqa
%

%=========

\begin{figure}[p] %[tbp]
	\centering
		\includegraphics[width=0.47
	\linewidth]{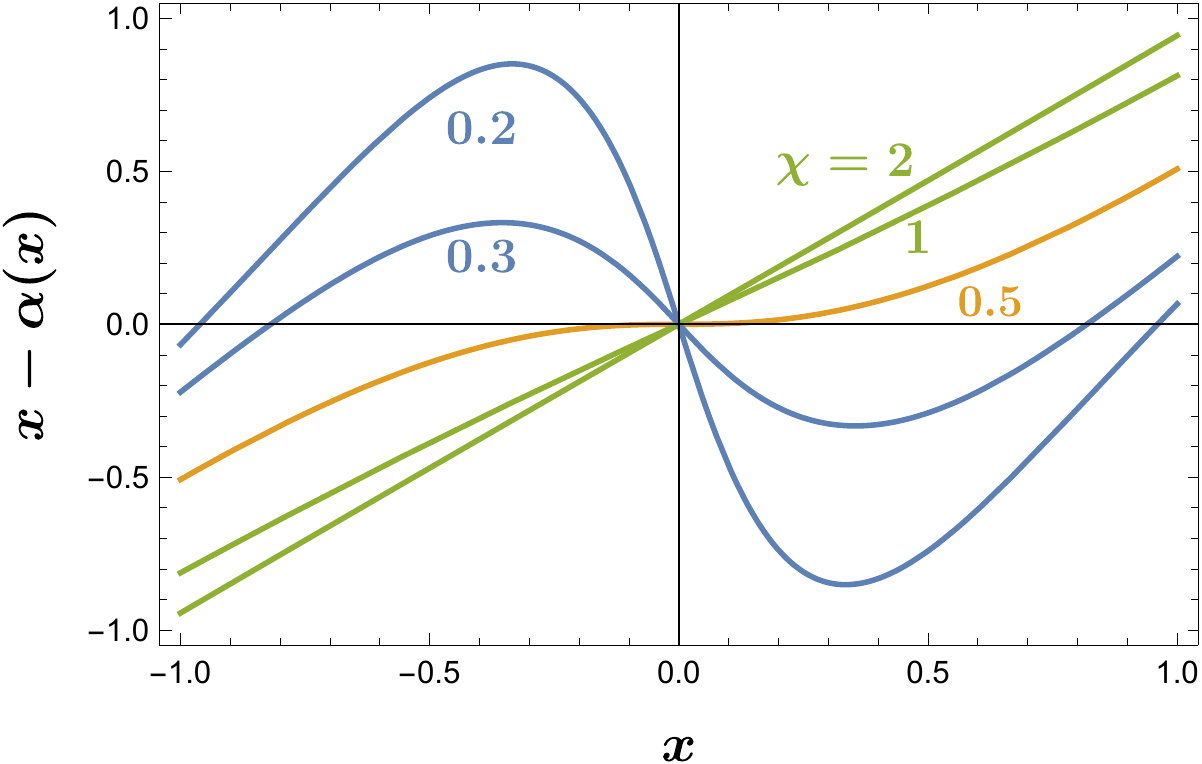} \quad \hspace{3mm}
	\includegraphics[width=0.45\linewidth]{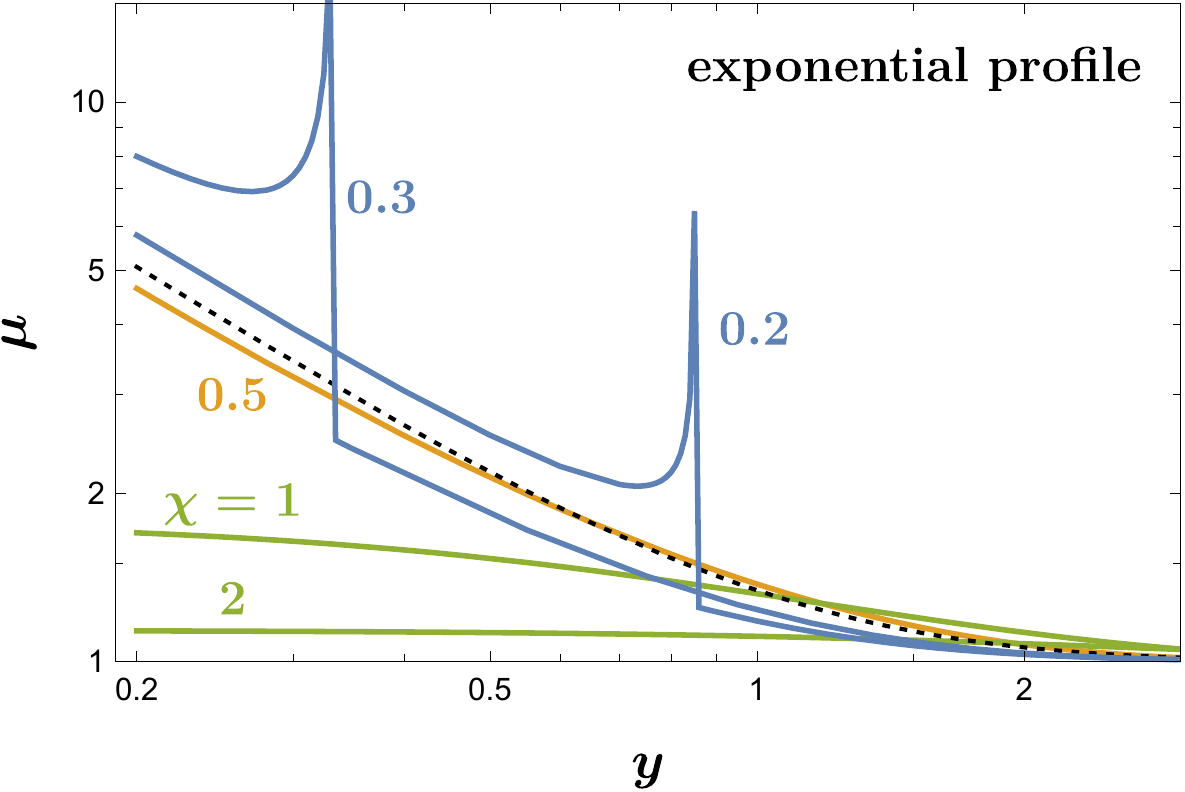} 
	\caption{The same as \fref{fig:xalpha-hard} but for dMACHO with an exponential mass density profile. }
	\label{fig:xalpha-exponential}
\end{figure}

%=========
Let us next consider a dMACHO whose mass density follows an exponential profile, $\rho(\vec{r})=\rho_0\,e^{-|\vec{r}|/R}$ with $\rho_0 = M/(8 \pi R^3)$, similar to the $1S$ radial wavefunction of the hydrogen atom.  
The surface density is calculated to be $\Sigma(x)=2\,\rho_0\,x\,K_1(x/\chi) \RE$ where $K_1(z)$ is the modified Bessel function.  
The dimensionless mass function $m(x)$ can then be expressed in terms of the generalized Meijer $G$ function as~\cite{generalized-Meijer-function} 
\beqa
m(x)^{\rm expon}=\frac{|x|^3}{8\,\chi^3}G^{2\,1}_{1\,3}\left(\dfrac{|x|}{2\, \chi},\dfrac{1}{2}\left| 
	\begin{matrix}%{ccc}
	\multicolumn{3}{c}{-1/2}\\
	-1/2 & 1/2 & -3/2\\
	\end{matrix}
\right. \right)\,.
\eeqa
The behavior of the function $x-m(x)/x$ in \eref{eq:lens} is similar to the uniform density model, except that the critical radius for having one or three different images is now $\chi=1/2$.  
In \fref{fig:xalpha-exponential} we present the magnification factor for different $\chi=R/\RE$.  
For a large radius with $\chi > 1/2$ and in the limit of $y \rightarrow 0$, the maximal magnification factor becomes 
\beqa\label{eq:mu_max_2}
\mu_\mathrm{max}^{\rm expon} \approx  \left(1 - \dfrac{1}{4\,\chi^2}\right)^{-2} ~.
\eeqa
%

%=========

%=========
Let us now address how a telescope may be sensitive to a dMACHO with an extended mass distribution.  
Suppose that a certain telescope observation is sensitive to a lensing event if the magnification factor is larger than some threshold value, $\mu \geq \mu_\mathrm{T}$; we can think of $\mu_\mathrm{T}$ as a measure of the telescope's sensitivity.  
We saw from \erefs{eq:mu_max_1}{eq:mu_max_2} that for large dMACHOs, $\chi = O(1)$, the magnification factor is bounded from above, corresponding to a lensing event with source along the lens axis.  
By inverting these relations, we can derive a necessary condition for the detection of a lensing event, namely that the dMACHO's size is below a threshold, $\chi < \chi_\mathrm{T}$, with 
\begin{align}
\label{eq:rT-hard-sphere}
\chi_\mathrm{T}^{\rm uniform} 
& = \sqrt{\dfrac{3}{2}}\, \sqrt{1 + \frac{1}{\sqrt{\mu_\mathrm{T}} - 1} } ~, \\
\label{eq:rT-expon}
\chi_\mathrm{T}^{\rm expon}  & = \dfrac{1}{2}\, \sqrt{1 + \frac{1}{\sqrt{\mu_\mathrm{T}} - 1} } \per
\end{align}
Recall that $\chi = R / \RE$ so an upper limit on $\chi$ implies an upper limit on $R / \sqrt{M}$ for a given system ($D_\mathrm{OL}$ and $D_\mathrm{OS}$).  
In principle different telescopes may set different sensitivity thresholds, $\mu_\mathrm{T}$, but it is conventional to choose $\mu_\mathrm{T} = 3 / \sqrt{5} \simeq 1.34$, since this corresponds to the value of $\mu$ for a point mass \pref{eq:mu-point-mass} at $y=1$.  
For $\mu_\mathrm{T}=1.34$ we evaluate $\chi_\mathrm{T}^{\rm uniform} \simeq 3.32$ and $\chi_\mathrm{T}^{\rm expon} \simeq 1.36$.   

\begin{figure}[t!] %[tbp]
	\centering
		\includegraphics[width=0.6
	\linewidth]{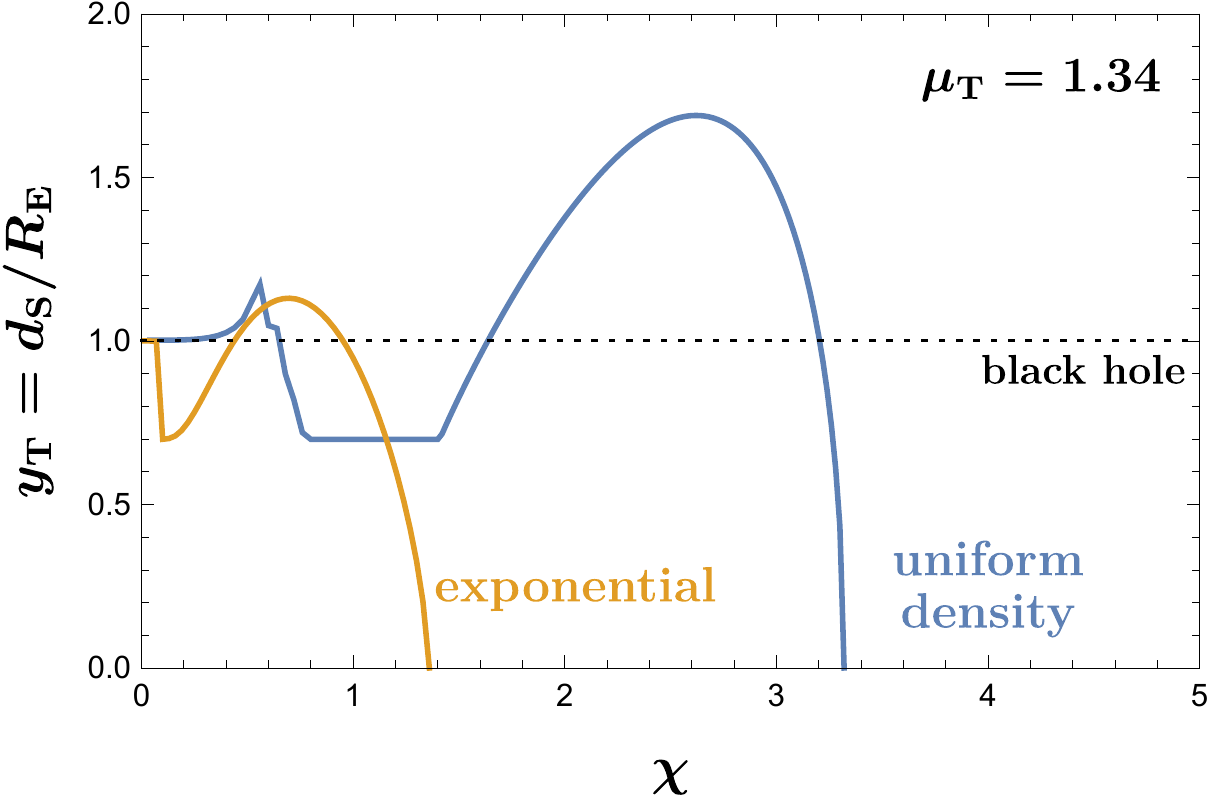}
	\caption{This figure shows the threshold value of the dimensionless source position on the lens plane, $y_{\rm T}$, as a function of the dMACHO radius, $\chi = R / \RE$.  If a given experiment can detect sufficiently large magnification factors $\mu \geq \mu_{\rm T} = 1.34$ then it can infer the presence of a dMACHO with a sufficiently small value of $y \leq y_{\rm T}$.  We show two dMACHO models, corresponding to a uniform density and an exponential density profile, and we compare with the case of a point-like black hole for which $y_{\rm T} \approx 1$ and there is no analog of $\chi = R / \RE$.  
	\label{fig:yT-r}
	}
\end{figure}

%=========
It is also useful to think of the lower limit on $\mu$ as an upper limit on the dimensionless source position, $y$.  
From the right panels of Figs.~\ref{fig:xalpha-hard}~and~\ref{fig:xalpha-exponential} we see that $\mu > \mu_\mathrm{T}$ will only be solved for sufficiently small $y$ at a given $\chi$.  
In \fref{fig:yT-r}, we show this threshold value of $y_\mathrm{T}$ as a function of $\chi$ for a fixed $\mu_\mathrm{T} = 1.34$.  
We find similar behavior for dMACHOs with either a uniform density or an exponential mass density profile.  
Note that $y_\mathrm{T} \to 0$ as $\chi \to \chi_{\rm T}$ given by \erefs{eq:rT-hard-sphere}{eq:rT-expon}, which reflects the fact that there are no solutions to $\mu > \mu_{\rm T}$ for $\chi > \chi_{\rm T}$.  

In addition to the uniform sphere and exponential mass profiles considered above, there also exist several other dMACHO mass profiles, \textit{e.g.} the power-law profile for self-similar subhalo~\cite{Bertschinger:1985pd}. For the following calculations we use the uniform sphere profile as a benchmark, and we refer the readers to~\cite{Croon:2020wpr} for the calculations and constraints of several other models.

%=========

%==================================
% Optical depth and microlensing constraints
%==================================
\subsection{Optical depth and microlensing constraints}\label{sub:lens_optical}

%=========
We now address how observational constraints on microlensing events can be used to place limits on dMACHOs in the Milky Way halo.  
In particular, we seek to recast existing limits on PBH dark matter.  
To do so, we will define an ``optical depth'' parameter, $\tau$.  
For PBHs we calculate $\tau$ as a function of $M$ and the dark matter fraction $f_\DM$ (since $R = \RS \ll \RE$ is not variable), and for dMACHOs we calculate it as a function of $M$ and $R$, assuming $f_\DM = 1$.  
Thus using $\tau$ as a bridge, we recast PBH limits in the $(M, f_\DM)$ plane into dMACHO limits in the $(M,R)$ plane.  

%=========
For a given source star, the ``optical depth'' $\tau$ counts the average number of lensing masses that reside within the lensing tube; \textit{i.e.} the axially-symmetric region of space between the observer and the source that has distance-dependent radius $y_\mathrm{T} \RE$.  
We can write the optical depth as 
\beqa
\tau = D_\mathrm{OS} \, \int^1_0 \! \ud\kappa \ n_\mathrm{lens}(\vec{r}_{\rm O} + \kappa \, D_{\rm OS} \, \hat{n}_{\rm L}) \ \pi \, y_\mathrm{T}^2\Big(R / \RE(\kappa), \mu_\mathrm{T}\Big)\,\RE^2(\kappa)  ~,
\eeqa
where $D_\mathrm{OS}$ is the distance from the observer to the source, $n_\mathrm{lens}(\vec{r})$ is the number density of lenses at location $\vec{r}$, $\vec{r}_{\rm O}$ is the location of the observer, $\hat{n}_{\rm L}$ is a unit vector pointing toward the source, $y_\mathrm{T}$ is shown in \fref{fig:yT-r} for $\mu_\mathrm{T} = 1.34$, and $\RE(\kappa)$ is the Einstein radius \pref{eq:Einstein-radius} of a lens with mass $M$ that's located at a distance $D_\mathrm{OL} = \kappa \, D_\mathrm{OS}$ from the observer. We suppose that the lenses can either be PBHs or dMACHOs. For PBHs we write $n_\mathrm{lens}(\vec{r}) = f_\PBH \ \rho_\DM(\vec{r}) / M$ where $\rho_\DM(\vec{r})$ is the mass density of dark matter at location $\vec{r}$, and where all PBHs are assumed to have a common mass $M$, and where $0 \leq f_\PBH \leq 1$ is the fraction of dark matter in the form of PBHs.  
For dMACHOs we write instead $n_\mathrm{lens}(\vec{r}) = \rho_\DM(\vec{r}) / M$ where we assume that dMACHOs make up all of the dark matter and that all dMACHOs have a common mass $M$.  
Since $n_\mathrm{lens} \propto 1/M$ and $\RE^2 \propto M$, we see that $\tau$ is dependent of $M$ for the PBH case (this is not true if the source size is large, see Refs.~\cite{Griest:2011av,Griest:2013aaa}), but $\tau$ depends on $M$ for the dMACHO case, because of the additional $M$ dependence in $y_\mathrm{T}$.  

%=========
We consider source stars in the Milky Way (MW) galaxy, the Large Magellanic Cloud (LMC), and the Andromeda galaxy (M31).  
To describe the dark matter distribution in these systems, we adopt the parametrization used by the MACHO/EROS experiments~\cite{Alcock:2000ph,Tisserand:2006zx}.  
In particular, we assume a spherically symmetric dark matter halo with mass density $\rho_\DM(r)$ at location $r = |\vec{r}|$, which is given by 
\beqa
\rho_\mathrm{DM}(r)= \rho_0 \,\frac{d_\odot^2 + a^2}{r^2 + a^2} \per
\eeqa
For the MW dark halo, $\rho_0 \approx 0.3 \GeV / \mathrm{cm}^3 = 0.0079 \Msun / \mathrm{pc}^3$ is the dark matter density at the Sun, $d_\odot = 8.5 \kpc$ is the Galactocentric radius of the Sun, and $a=5 \kpc$.  
For the LMC dark halo, Refs.~\cite{Alcock:2000ph,Tisserand:2006zx} take $\rho_0 = 0.0223 \Msun / \mathrm{pc}^3$ and $a=2 \kpc$.  
As shown in Ref.~\cite{Alcock:2000ph}, the LMC halo only provides a small contribution to the total optical depth, and therefore we will include only the MW dark halo's contribution to calculate $\tau$.  
Note that a source in the LMC is a distance $D_\mathrm{OS} \approx 50 \kpc$ away from Earth, and the LMC is located at $(\ell, b) = (280.47^\circ, -32.75^\circ)$ in galactic coordinates~\cite{2014ApJ...792...43F}.  
%\SL{Comment on the DM density profile choice.}
The limits derived by Subaru/HSC and EROS/MACHO assume NFW and isothermal dark matter density profiles in the Milky Way halo, respectively.  
If dark matter has an appreciable clustered component, the limit can be changed, which has been studied in the context of PBH dark matter in Ref.~\cite{Garcia-Bellido:2017xvr}.  

\begin{figure}[t!] %[tbp]
	\centering
		\includegraphics[width=0.6
	\linewidth]{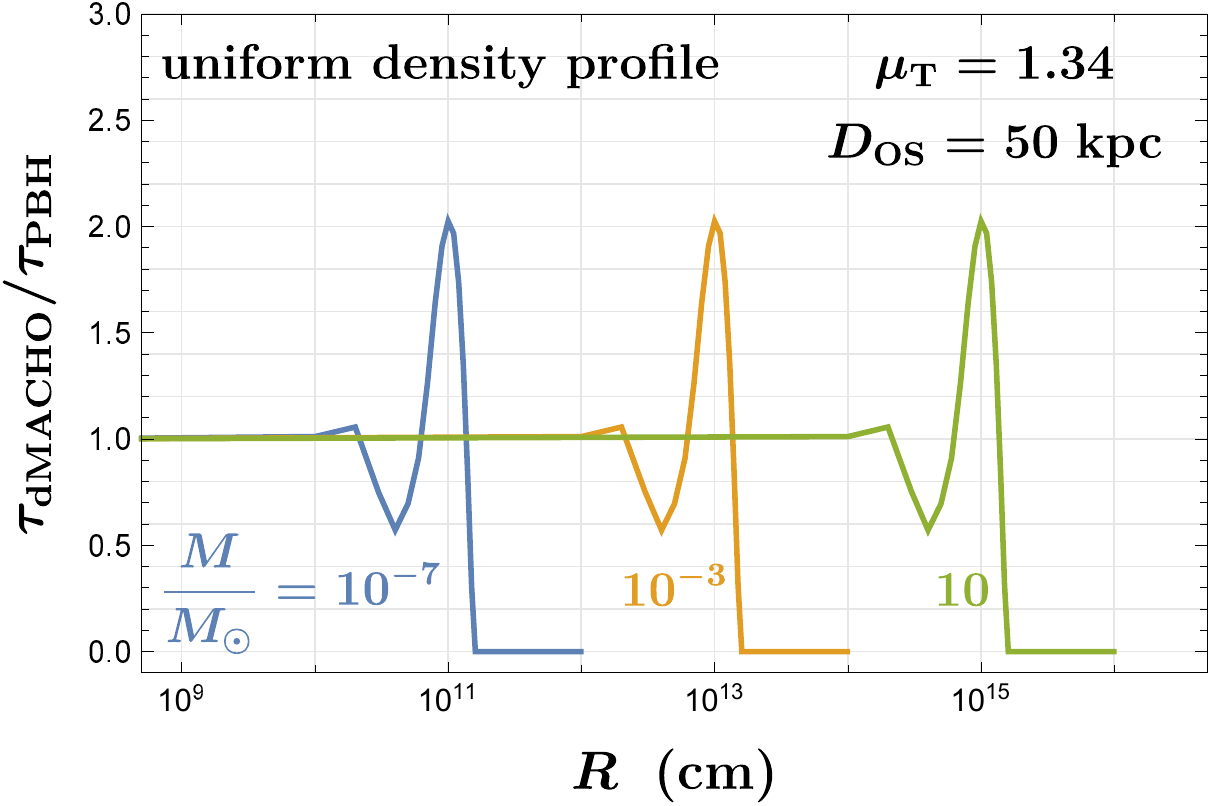}
	\caption{The ratio of the optical depth for dMACHO over the black hole case. The source location is chosen to be LMC with a distance to the Earth around 50 kpc.  }
	\label{fig:optical-depth-ratio}
\end{figure}
%

%=========
Using the formulas above, we calculate $\tau$ for both the PBH and dMACHO cases.  
For the PBH case we find $\tau_\PBH \approx 4.7\times 10^{-7}$, consistent with earlier results in Ref.~\cite{Paczynski:1985jf}, which is independent of $M$ as anticipated.  
For the dMACHO case, we take into account the radius-dependent $y_\mathrm{T}$ as shown in \fref{fig:yT-r} and calculate $\tau_\mathrm{dMACHO}$ for different masses and radii.  
The predicted optical depths are summarized in \fref{fig:optical-depth-ratio}.
For a small dMACHO, $R \ll \RE$, the dMACHO's optical depth asymptotes to the same value as a PBH of equal mass.  
For a larger dMACHO with $R \gtrsim \RE$, the optical depth drops very quickly.  
Thus there is a threshold dMACHO radius, $R \sim \RE$, such that larger dMACHOs are not probed by EROS/MACHO.  
In these relations, ``$\RE$'' corresponds to \eref{eq:Einstein-radius} with $\kappa=1/2$.  The dip and bump structure in this plot comes from the dependence of threshold dimensionless source distances on $R$, as can been seen from Fig.~\ref{fig:yT-r}.

\begin{figure}[t!] %[tbp]
	\centering
		\includegraphics[width=0.6
	\linewidth]{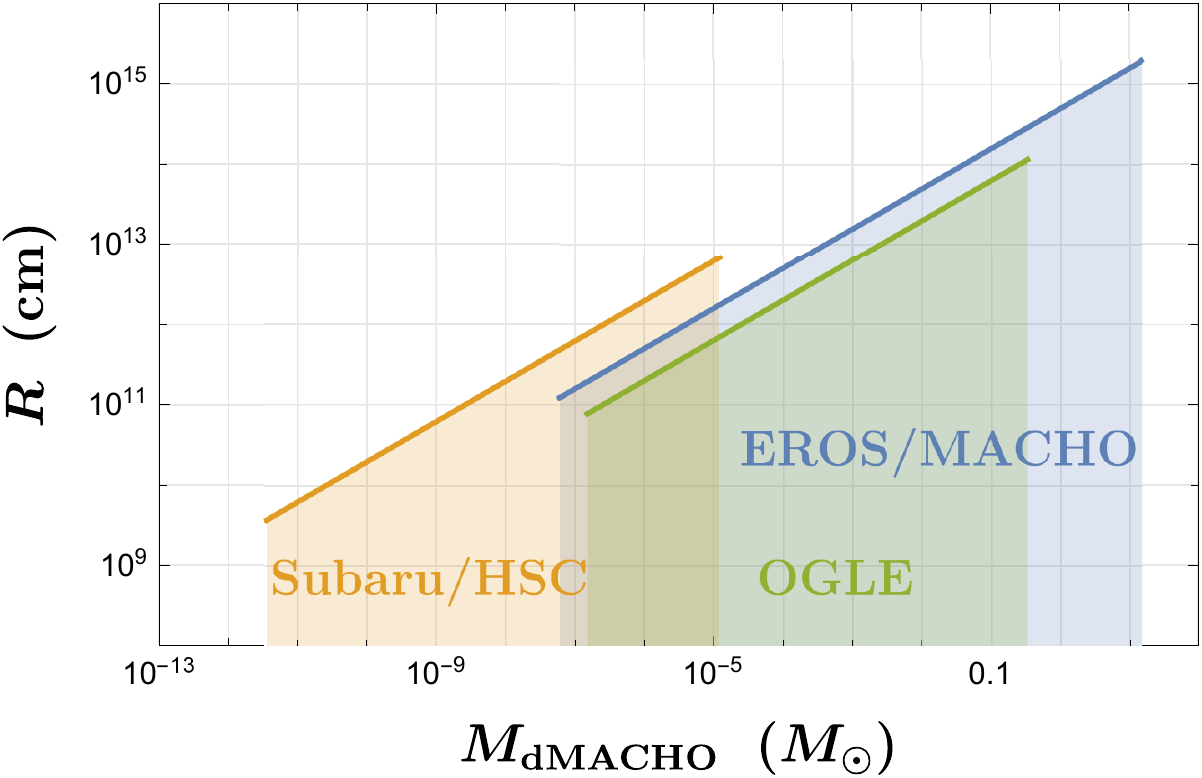}
	\caption{
	The constraints on dMACHO radii as a function of dMACHO mass based on microlensing experiments, assuming that dMACHO contributes 100\% of dark matter.  
The uniform density profile is used here, 
while the exponential profile has a smaller upper limit by a factor of around 2.4. }
	\label{fig:lensing-recast}
\end{figure}

%=========
Now we are equipped to recast the PBH lensing limits onto dMACHOs.  
Numerically, the threshold radius probed by a lensing experiment is very close to $\chi_\mathrm{T}^{\rm uniform}$ in \eref{eq:rT-hard-sphere} for the uniform density profile and $\chi_\mathrm{T}^{\rm expon}$ in \eref{eq:rT-expon} for the exponential profile. 
The reason behind this can be seen from the features of Fig.~\ref{fig:optical-depth-ratio}. 
For a given mass, the constraints on $f_\PBH$ for the PBH case mean $\tau_\mathrm{dMACHO}/\tau_\mathrm{PBH} < f^\mathrm{max}_\PBH$ assuming that dMACHO accounts for 100\% of dark matter. 
%\SL{Needs rewording.} 
%[For a value of $f^\mathrm{max}_\PBH$ below but not close to 1, the intersecting values of $R$ from the curves in Fig.~\ref{fig:optical-depth-ratio} are very close to the end points of the curves or values in \eref{eq:rT-hard-sphere}.] 
The value of $R$ at which $\tau_\mathrm{dMACHO} / \tau_\mathrm{PBH}$ drops to zero is well approximated by the $R = \chi_\mathrm{T} \, R_\mathrm{E}$ with $\chi_\mathrm{T}$ given by \eref{eq:rT-hard-sphere}.
All the experiments we consider, EROS/MACHO, Subaru/HSC and OGLE, choose the threshold magnification to be $\mu_\mathrm{T}$ =1.34, and one therefore has $\chi_\mathrm{T}^{\rm uniform}\approx 3.32$ and $\chi_\mathrm{T}^{\rm expon} \approx 1.36$. 
So, for the uniform density profile, the constraints on dMACHO radii from EROS/MACHO~\cite{Tisserand:2006zx} with $D_\mathrm{OS} \approx 50 \kpc$ for LMC can be recast into
\beqa
R\gtrsim (5.0 \times 10^{14}\,\mbox{cm}) \, \left(M/M_\odot\right)^{1/2} \,, \qquad 
0.6\times 10^{-7} \, M_\odot < M < 15\, M_\odot ~ \mbox{[EROS/MACHO]}~, 
\eeqa
with a factor of 2.4 smaller for the exponential profile. We note that once a PBH mass is constrained by a certain microlensing experiment, the constraints on dMACHO radius will be approximately given by the above formula and insensitive to how small $f_\PBH$ is constrained. For the Subaru/HSC~\cite{Niikura:2017zjd}, M31 has been used as the source location with $D_\mathrm{OS} \approx 770 \kpc$. For OGLE~\cite{Niikura:2019kqi}, the sources in the MW bulge have been used with roughly the average distance to be around $8 \kpc$. The recast limits on $R$ from them are
\beqa
&& R\gtrsim (2.0 \times 10^{15}\,\mbox{cm}) \, \left(M/M_\odot\right)^{1/2} \,, \qquad 
3.6\times 10^{-12} \, M_\odot < M < 1.2\times 10^{-5}\, M_\odot ~ \mbox{[Subaru/HSC]}~, \nonumber \\
 && R\gtrsim (2.0 \times 10^{14}\,\mbox{cm}) \, \left(M/M_\odot\right)^{1/2} \,, \qquad 
1.5\times 10^{-7} \, M_\odot < M < 0.3\, M_\odot ~ \mbox{[OGLE]}~. 
\eeqa
In \fref{fig:lensing-recast}, we summarize the constraints on the dMACHO radius based on the three most-constraining microlensing experiments.  
In this figure, we have assumed that 100\% of dark matter is made of dMACHOs.  
If the dMACHO is only a fraction of dark matter, the constrained range of dMACHO mass from each experiment shrinks and approximately matches the corresponding range for PBH with the same fraction. 

%=========
%
%

%=========
Before we end this section, let us discuss how future observations could be used to probe even larger dMACHOs.  
Recall that the lower limit on the dMACHO's scale radius, $R > \chi_\mathrm{T} \RE$, can be strengthened if $\RE$ is increased or if $\mu_\mathrm{T}$ is decreased, which would increase $\chi_\mathrm{T}$ via \erefs{eq:rT-hard-sphere}{eq:rT-expon}.  
To increase $\RE$ through \eref{eq:Einstein-radius}, one could look for a system that's further away from Earth, but this comes at the cost of a reduction in the source flux, making it harder to measure the light curve.  
To decrease $\mu_\mathrm{T}$, one can look for telescopes with exceptional photometric precision~\cite{Griest:2011av,Ricker_2014}; taking $\mu_\mathrm{T}-1$ as small as $10^{-3}$ increases the upper limit on $R$ by a factor of 17 compared to $\mu_\mathrm{T} =1.34$.

%==================================
% Accretion of baryonic matter
%==================================
\section{Accretion of baryonic matter}\label{sec:accretion}

%=========
In the previous section we have studied the effect of a dMACHO's gravitational influence on light through the phenomenon of gravitational lensing, and in this section we turn our attention to the dMACHO's influence on matter.  
As gravitational sources, dMACHOs can accumulate matter inside and around themselves, and since the accreted matter is hotter than the surrounding medium, the dMACHOs will develop a ``glowing'' halo of baryonic matter.  
In this section we first calculate the density, temperature, and ionization profiles of the accreted baryonic matter; we next use these profiles to calculate the spectrum and luminosity of the glowing halo; and we finally study cosmological and astrophysical probes of glowing dMACHOs.  
For quantitative calculations we assume the uniform dMACHO 
density profile \pref{eq:hard_sphere}, but we anticipate qualitatively similar results for other profiles, {\it e.g.} the exponential density profile discussed  in the previous section.  
In our study, we only consider the spherically symmetric accretion, although we note that the non-spherically symmetric accretion could provide stronger signals, as with black holes~\cite{Poulin:2017bwe}. 

%==================================
% Spherical accretion onto an extended mass
%==================================
\subsection{Spherical accretion onto an extended mass}\label{sec:accretion-equations}

%=========
In this subsection we study the accretion of matter onto a dMACHO.  
We will see that the dMACHO's attractive gravitational force may cause it to develop a ``halo'' of baryonic matter, and we will calculate the density, temperature, and ionization fraction of this plasma.  
In the next subsection, we will study electromagnetic radiation from the dMACHO's accreted matter.  
The formalism used in this subsection will be familiar to readers acquainted with Bondi accretion~\cite{Bondi:1952ni}, which is often used to study accretion in astrophysical environments such as around stars and black holes~\cite{Shapiro_BH}.  

%=========
\subsubsection{Gravitational potential}
Consider a dMACHO with mass $M$, scale radius $R$, and a spherically-symmetric enclosed mass profile $\widetilde{M}(r)$ where $r = |\vec{r}|$ is the distance from the center of the dMACHO.  
The dMACHO exerts an attractive gravitational force on the surrounding matter, and the corresponding gravitational acceleration $\vec{g}_\mathrm{grav}(\vec{r})$ is written as 
\begin{align}\label{eq:g_grav}
	\vec{g}_\mathrm{grav}(\vec{r}) = g_\mathrm{grav}(r) \ \vec{r} / r ~, 
	\qquad \text{with} \qquad  
	g_\mathrm{grav}(r) = - \GN \widetilde{M}(r) \, / \, r^2 
	\per
\end{align}
We assume that the population of dMACHOs is dilute, and we study accretion onto an isolated dMACHO.~\footnote{This is an excellent approximation for the systems of interest, even if dMACHOs make up all of the dark matter.  The typical separation between dMACHOs is estimated as $l \sim \rho_\mathrm{dMACHO}^{-1/3} M^{1/3}$, where $\rho_\mathrm{dMACHO}$ is the mass density of the dMACHO population.  The assumption of isolated accretion is justified if $l \gg \RB$, where the Bondi radius, $\RB \propto M$ from \eref{eq:RB_and_tB}, sets the typical size of the accretion region.  The dMACHOs are diluted provided that $M < c_\infty^3 \GN^{-3/2} \rho_\mathrm{dMACHO}^{-1/2}$, which, for example, evaluates to $M < 1.7\times 10^5 \Msun$ at recombination when $c_\infty \simeq 2.2\times 10^{-5}$ and $\rho_\mathrm{dMACHO} = \rho_\mathrm{DM} = 1.3\times 10^{-38}\,{\rm GeV}^4$.
}  
We also ignore the contribution to the gravitational potential from the accreted baryonic matter, which is much smaller than the dMACHO mass.  

%=========
\subsubsection{Baryonic matter}
When the dMACHO is placed in a medium, it will accrete the surrounding matter due to its gravitational pull.  
For now we assume that the dMACHO is at rest with respect to the medium, and later in this section we will introduce a nonzero relative velocity between the two.  
We also assume that the medium is approximately homogeneous, which is a well-justified assumption for the primordial plasma.  
Consequently, this motivates us to assume that the entire accretion process will be spherically-symmetric.  
Namely, we assume that scalar quantities are only functions of time $t$ and the radial coordinate $r = |\vec{r}|$, measured from the dMACHO's center, while vector quantities are proportional to $\vec{r} / r$.  

%=========
We assume that the accreting matter consists of electrons, protons, and neutral hydrogen atoms ($\tensor[^1]{\mathrm{H}}{}$), and we refer to these three components collectively as the ``baryonic matter.''    
For the systems of interest, heavier nuclei are less abundant, and they can be neglected.  
Let $n_e(r,t)$, $n_p(r,t)$, and $n_H(r,t)$ be the number densities of electrons, protons, and hydrogen atoms, respectively.  
We assume local charge neutrality, which imposes the constraint 
\beq\label{eq:nQ_def}
	n_e = n_p
	\per
\eeq
Let $\rho(r,t)$ be the mass density of the baryonic matter, which is given by 
\beq\label{eq:rho_def}
	\rho = m_e \, n_e + m_p \, n_p + m_H \, n_H 
	\com
\eeq
where $m_e \simeq 0.511 \MeV$ and $m_p \approx m_H \simeq 0.938 \GeV$ are the masses of an electron, proton, and hydrogen atom, respectively.  
Let $x_e(r,t)$ be the ionization fraction, which is given by 
\begin{align}\label{eq:xe_def}
	x_e = \frac{n_e}{n_e + n_H} 
	\per
\end{align}
Combining Eqs.~(\ref{eq:nQ_def}),~(\ref{eq:rho_def}),~and~(\ref{eq:xe_def}) gives 
\begin{align}\label{eq:ne_np}
	n_e = n_p \approx \frac{1}{m_p} \, x_e \, \rho ~,
	\qquad \text{and} \qquad 
	n_H \approx \frac{1}{m_p} \, \bigl( 1 - x_e \bigr) \, \rho 
	\com
\end{align}
where we have used $m_e \ll m_p \approx m_H$, but we have not assumed $x_e \ll 1$.  

%=========
\subsubsection{Local thermal equilibrium}
We assume that the three components of the baryonic matter are kept in local thermal equilibrium at a common temperature $T(r,t)$.  
For the temperature range of interest, the protons and hydrogen atoms are always cold ($T \ll m_p \approx m_H$), while the electrons may be either hot ($T \gg m_e$) or cold ($T \ll m_e$).  
Using Fermi-Dirac statistics for electrons and Maxwell-Boltzmann statistics for protons and hydrogen, we can evaluate the internal kinetic energy density and pressure of the baryonic matter, $\Ecal(r,t)$ and $P(r,t)$, which are found to be~\footnote{For cold species ($T/m \ll 1$) these relations are insensitive to the chemical potentials.  For hot species ($T/m \gg 1$) these relations hold for $\mu/T \ll 1$, which is the regime of interest.}  
\begin{align}
	\Ecal & = \frac{3}{2} \, T\,\big[n_e \, f_\Ecal(T/m_e) +  n_p + n_H\big] ~, \\ 
	P & = T\,\big[n_e \, f_P(T/m_e) +  n_p +  n_H \big] ~,
\end{align}
where $f_\Ecal(X)$ interpolates from $f_\Ecal(X \ll 1) \approx 1$ to $f_\Ecal(X \gg 1) \approx 7 \pi^4 / [270 \zeta(3)] \simeq 2.10$
and $f_P(X)$ interpolates from $f_P(X \ll 1) \approx 1$ to $f_P(X \gg 1) \approx 7\pi^4 / [540 \zeta(3)] \simeq 1.05$.  
By further using \eref{eq:ne_np}, we can write these equations as 
\begin{align}
	\Ecal & = \frac{3}{2} \, \frac{1}{m_p} \, T \, \rho \, \Bigl[ 1 + x_e \, f_\Ecal(T/m_e) \Bigr] \label{eq:E_from_rho} ~, \\ 
	P & = \frac{1}{m_p} \, T \, \rho \Bigl[ 1 + x_e \, f_P(T/m_e) \Bigr] \label{eq:P_from_rho}
	\per
\end{align}
It is useful to define the adiabatic index $\gamma(r,t)$, which is given by 
\begin{align}\label{eq:gamma_def}
	\gamma \equiv 1 + P / \Ecal
	= \bar{\gamma} \big( T(r,t)/m_e \big)
	\qquad \text{with} \qquad 
	\bar{\gamma}(X) \equiv 1 + \frac{2}{3} \ \frac{1 + x_e \, f_P(X)}{1 + x_e \, f_\Ecal(X)} 
	\per
\end{align}
For $x_e \ll 1$ we have $\gamma \approx 5/3$, independent of $T/m_e$, while for $x_e \approx 1$ we have $\gamma \approx 5/3$ for $T/m_e \ll 1$ and $\gamma \approx 13/9$ for $T/m_e \gg 1$.~\footnote{For practical applications, we remark that $f_\Ecal$ and $\bar{\gamma}$ are well approximated by empirical fitting formulas, $f_\Ecal(X) \simeq 1+(1.1\,X)/(X+0.78)$ and $\gamma(X) = (5/3) - (2/9) X/(X+0.47)$ for $x_e\approx 1$.}

%=========
\subsubsection{Fluid equations}
The mass density $\rho(r,t)$, fluid velocity $\vec{v}(\vec{r},t) = v(r,t)\,\vec{r}/r$, and internal energy density $\Ecal(r,t)$ are related by a system of first-order differential equations that encode the conservation of mass, momentum, and energy.  
These Navier-Stokes equations are written as~\cite{draine2010physics}
\begin{subequations}\label{eq:fluid_eqns}
\begin{align}
\label{eq:mass}
	& \dot{\rho} + \frac{1}{r^2} \bigl( r^2 \rho v \bigr)^\prime = 0 \ , \\ 
\label{eq:momentum}
	&\rho \, \dot{v} + \rho \, v \, v^\prime + P^\prime = \rho g \ , \\ 
\label{eq:energy}
	& \rho \bigl( \Ecal / \rho \bigr)^{\bullet } + \rho \, v \, \bigl( \Ecal / \rho \bigr)^\prime + P \frac{1}{r^2} \bigl( r^2 v \bigr)^\prime = \dot{q} \ , 
\end{align}
\end{subequations}
where $\dot{\chi} = \partial \chi / \partial t$ and $\chi^\prime = \partial \chi / \partial r$.  
In the second equation, $\vec{g}(\vec{r},t) = g(r,t) \, \vec{r} / r$ with $g = g_\mathrm{grav} + g_\mathrm{drag}$ represents the acceleration induced by the dMACHO's gravitational influence \pref{eq:g_grav} and by additional sources of drag.  
In the third equation, $\dot{q}(r,t)$ represents the rate of heating (if $\dot{q} > 0$) or cooling (if $\dot{q} < 0$) per unit volume.  
Interactions between the accreting matter and the CMB radiation can lead to nonzero drag and cooling~\cite{peebles1980large}
\begin{align}
	g_\mathrm{drag} & = - \frac{4}{3} \, \frac{x_e\, \sigma_\mathrm{T}\, \rho_\mathrm{cmb}}{m_p} \, v \label{eq:gdrag_def}  \com\\ 
	\dot{q} & = \frac{4\,x_e \, \sigma_\mathrm{T}\, \rho_\mathrm{cmb}}{m_p m_e} \, \bigl( T_\mathrm{cmb} - T \bigr) \rho \label{eq:qdot_def}
	\com 	
\end{align}
where $\sigma_\mathrm{T} \simeq 6.65 \times 10^{-25} \cm^2$ is the Thomson scattering cross section, $\rho_\mathrm{cmb} \simeq (1.97 \times 10^{-15} \eV^4) (1+z)^4$ is the CMB energy density, and $T_\mathrm{cmb} \simeq (2.34 \times 10^{-4} \eV) (1+z)$ is the CMB temperature at redshift $z$.  
Upon substituting \erefs{eq:E_from_rho}{eq:P_from_rho} into \eref{eq:fluid_eqns}, the dynamical variables become $\rho(r,t)$, $v(r,t)$, and $T(r,t)$.  
A separate equation that determines $x_e(r,t)$ will be discussed below.  

%=========
\subsubsection{Simplifying assumptions}
Here we make several simplifying assumptions that allow us to solve the fluid equations.   
\begin{enumerate}
%---
	\item  We are not interested in the dynamical process of accretion, but only in the stationary configuration that results after accretion is completed.  The stationary configuration is a solution of \eref{eq:fluid_eqns} with $\partial / \partial t\,\to\,0$.  The profile functions for the stationary solutions are written as $\rho(r)$, $v(r)$, $T(r)$, $\Ecal(r)$, $P(r)$, $x_e(r)$, and $\gamma(r)$.  
%---
	\item  We do not attempt to solve the fluid equations for general $\gamma(r)$, but instead we now suppose that $\gamma(r)$ is a constant.  We will solve \eref{eq:fluid_eqns} separately in the outer region where $T \ll m_e$ and $\gamma(r) \approx 5/3$ and in the inner region where $T \gg m_e$ and $\gamma(r) \approx 13/9$, and we will construct the full profile by matching these solutions at $T \sim m_e$.  
%---
	\item  If the temperature of the accreted matter exceeds $T \sim 1 \eV$, then the neutral hydrogen will be ionized.  This means that the ionization fraction will vary from $x_e = \bar{x}_e$ at large $r$ to a value as large as $x_e = 1$ at small $r$.  
For the present discussion we will assume $x_e(r)$ to be constant, which is a good approximation away from the region where the ionizing phase transition takes place, and we will discuss below how to account for the ionization.  
%---
	\item  We first assume that the drag force and cooling rate, parametrized by $g_\mathrm{drag}$ and $\dot{q}$ respectively, can be neglected.  We will argue below that $g_\mathrm{drag}$ is generally negligible for the systems of interest. The cooling rate $\dot{q}$ will be taken into account in our study, although its influence on our CMB anisotropy calculations is also not important. 	
%---
	\item  For the stationary solution described above, we expect that the flow speed will vanish, and therefore we implement the {\it hydrostatic approximation} by taking $v(r) = 0$.  The correct physical picture for the stationary solution is one in which the pressure exerted by the accreted matter balances the dMACHO's gravitational force, and no further matter is accreted.  Then \eref{eq:momentum} is reduced to simply $P^\prime = \rho \, g_\mathrm{grav}$, while \erefs{eq:mass}{eq:energy} are satisfied trivially.  Note that the physical picture here is different from the one that's often used to study accretion onto black holes.  We discuss the distinction between hydrostatic accretion and Bondi accretion below.  
%---
	\item  We assume a polytropic equation of state with a constant adiabatic index $\gamma$.  This relates the pressure and mass density according to~\footnote{It is possible to derive this relation from \eref{eq:fluid_eqns} in the stationary approximation without heating.  Combining \erefs{eq:mass}{eq:energy} gives $\Ecal \bigl[ \dot{\Ecal} / \Ecal - \gamma \dot{\rho} / \rho \bigr] + \Ecal \vec{v} \cdot \bigl[ \vec{\nabla} \Ecal / \Ecal - \gamma \vec{\nabla} \rho/\rho \bigr] = \dot{q}$ where $\gamma = 1 + P/\Ecal$ is given by \eref{eq:gamma_def}.  If $\dot{\Ecal} = \dot{\rho} = \dot{q} = 0$ and if $\gamma$ is a constant, then the solution is $\Ecal = C \rho^\gamma$ or equivalently $P = K \rho^\gamma$.  }
\begin{align}\label{eq:P}
	P(r) = K \, \rho(r)^{\gamma} ~,
\end{align}
where $K$ is a multiplicative constant with mass dimension $4-\gamma$, and $\gamma$ is the adiabatic index from \eref{eq:gamma_def}.  
Using \eref{eq:P_from_rho} it follows that 
\begin{align}\label{eq:T}
	T(r) = K \, m_p \, \frac{\rho(r)^{\gamma-1} }{1+x_e \, f_P}
	\per
\end{align}
If the profile has a point where $T/m_e \approx 1$ (electrons become relativistic) then this identifies a boundary between two regions with different, constant values of $f_\Ecal$, $f_P$, $\gamma$, and $K$.  
%---
	\item  We are interested in solutions that obey the boundary conditions
\begin{align}\label{eq:BCs}
	\hspace{-2mm} \lim_{r \to \infty} \rho(r) \, \to \, \rho_\infty
	\, , \hspace{5.5mm} 
	\lim_{r \to \infty} T(r) \, \to \, T_\infty
	\, , \hspace{5.5mm} 
	\lim_{r \to \infty} x_e(r) \, \to \, \bar{x}_e
	\com 
\end{align}
imposed at spatial infinity.  Then Eqs.~(\ref{eq:E_from_rho}),~(\ref{eq:P_from_rho}),~and~(\ref{eq:gamma_def}) define $\Ecal_\infty$, $P_\infty$, and $\gamma_\infty$.  For the systems of interest, we have $T_\infty / m_e \ll 1$ and so 
\begin{align}
	\Ecal_\infty & = \frac{3}{2} \, \frac{1}{m_p} \, T_\infty \, \rho_\infty \, \bigl[ 1 + \bar{x}_e \bigr] \label{eq:E_infty} \com \\ 
	P_\infty & = \frac{1}{m_p} \, T_\infty \, \rho_\infty \bigl[ 1 + \bar{x}_e \bigr] \label{eq:P_infty}  \com\\ 
	\gamma_\infty & = 1 + P_\infty / \Ecal_\infty = 5/3 \label{eq:gamma_infty}
	\per
\end{align}
It is useful to define the Bondi radius $\RB$ and Bondi time $\tB$, which are given by 
\begin{align}\label{eq:RB_and_tB}
	\RB \equiv \GN M / c_\infty^2  
	\qquad \text{and}  \qquad
	\tB \equiv \GN M / c_\infty^3 
	\com
\end{align}
where $M$ is the total dMACHO mass and where 
\begin{align}\label{eq:c_infty}
	c_\infty \equiv \sqrt{ \gamma_\infty P_\infty / \rho_\infty} 
\end{align}
is the asymptotic adiabatic sound speed far away from the dMACHO.  
%---
	\item  For the parameters of interest, accretion occurs quickly compared to the cosmological time scale, \textit{i.e.} $\tB \ll \tH = H^{-1}$.  Due to the cosmological expansion, the ``stationary'' solution evolves adiabatically on times scales of $O(\tH)$, and when appropriate, we take this into account by including a time dependence in the boundary conditions \pref{eq:BCs}.  
\end{enumerate}
Under the assumptions specified here, the Navier-Stokes equation \pref{eq:momentum} reduces to 
\begin{align}\label{eq:drho_dr}
	& \frac{\GN \widetilde{M}(r)}{r^2} + \gamma \, K \, \rho(r)^{\gamma-2} \, \frac{d\rho}{dr} = 0 
	\per
\end{align} 
We solve this equation assuming that the dMACHO has a uniform-density mass profile with radius $R$, as in \eref{eq:hard_sphere}.  

%=========
\subsubsection{Outer adiabatic region}
Let us first consider the outermost region of the accretion volume, which extends out to the Bondi radius where $r = \RB$.  
For $r \gtrsim \RB$ the thermal velocity dispersion of the baryonic matter exceeds the dMACHO's escape velocity, and our accretion calculation does not apply.  
By imposing the boundary conditions in \eref{eq:BCs}, we solve \eref{eq:drho_dr} to find
\begin{subequations}\label{eq:outer}
\begin{align}
%\hspace{-10mm}
%---
\label{eq:rho_outer}
	\rho(r) = \rho_\infty \times \begin{cases}
	\left(1 + \dfrac{3(\gamma_\infty-1)}{2}\dfrac{\RB}{R} - \dfrac{\gamma_\infty-1}{2} \, \dfrac{r^2 \RB}{R^3} \right)^{\frac{1}{\gamma_\infty-1}} & , \ r \leq R \quad \text{(inside dMACHO)} ~
	\vspace{2mm}\\
	\left(1 + (\gamma_\infty-1) \dfrac{\RB}{r}\right)^{\frac{1}{\gamma_\infty-1}} & , \ R < r \lesssim \RB \quad \text{(outside dMACHO)} 
	\end{cases} ~,
\end{align}
where $\gamma_\infty = 5/3$.  
Here we have used \eref{eq:P} to write $K = P_\infty / \rho_\infty^{\gamma_\infty}$; we have used \eref{eq:c_infty} to write $P_\infty = c_\infty^2 \rho_\infty / \gamma_\infty$; and we have used \eref{eq:RB_and_tB} to write $\GN M = c_\infty^2 \RB$.  
The temperature profile is calculated using \eref{eq:T}, which gives 
\begin{align}
%---
\label{eq:T_outer}
	T(r) & = T_\infty \times \begin{cases}
	\left(1 + \dfrac{3(\gamma_\infty-1)}{2}\dfrac{\RB}{R} - \dfrac{\gamma_\infty-1}{2} \, \dfrac{r^2 \RB}{R^3} \right) & , \ r \leq R ~,
	\vspace{2mm}\\
	\left(1 + (\gamma_\infty-1) \dfrac{\RB}{r}\right) & , \ R < r \lesssim \RB 
	\end{cases} \per
\end{align}
\end{subequations}
In the regime where $R \ll r \ll \RB$ we find the following power law scaling behaviors:  $\rho \propto r^{-1/(\gamma_\infty-1)} = r^{-3/2}$ for the density and $T \propto r^{-1}$ for the temperature implying $\Ecal, P \propto r^{-\gamma_\infty/(\gamma_\infty-1)} = r^{-5/2}$ for the internal energy density and pressure.  
Note that the maximum temperature is obtained as $r \to 0$, which gives 
\begin{align}
	T_\mathrm{max} = T(r=0) \approx 
	 \frac{\RB}{R} \, T_\infty~,
\end{align}
where we have used $\gamma_\infty = 5/3$ and $\RB/R \gg 1$.  
If $T_\mathrm{max} < \mathrm{eV}$ then these solutions are valid for all $r$, but if $T_\mathrm{max} > \mathrm{eV}$ then it is necessary to take ionization into account, and if $T_\mathrm{max} > T_\mathrm{rel} \sim \MeV$ then it is necessary to take the relativistic electrons into account.  
We discuss these special regimes in the subsections below.  
The total accreted mass is calculated from $\rho(r)$ by performing the volume integral.  
For example if $T_\mathrm{max} < T_\mathrm{rel}$ then the amounts of accreted mass within the Bondi radius $\RB$ and within the dMACHO radius $R$ are found to be 
\begin{align}\label{eq:accrete-mass-RB}
	&M_\mathrm{accrete}^{r \leq \RB} \simeq 2.5 \, \pi\,\RB^3\,\rho_\infty = 2.5 \, \pi \, \GN^3\,M^3\, \dfrac{\rho_\infty}{c^6_\infty} ~,\\
	&M_\mathrm{accrete}^{r \leq R} \simeq 0.96\,\pi\,R^{3/2}\,\RB^{3/2}\,\rho_\infty = 0.96\, \pi  \,R^{3/2}\, \GN^{3/2}\,M^{3/2}\, \dfrac{\rho_\infty}{c^3_\infty}  ~,
\end{align}
where we have taken $\gamma_\infty = 5/3$ and kept only the leading terms in $R/\RB \ll 1$.  
For the systems of interest, the accreted mass is much smaller than the dMACHO mass $M$.  

The solution \eref{eq:outer} is valid only when the accretion profile has non-relativistic electrons, negligible cooling from CMB and the constant ionization region. In the following subsections we will discuss the modifications to this solution by including these ignored effects one by one and summarize the results.

%=========
\subsubsection{Inner adiabatic region}
If the accretion is very strong, then the plasma near the dMACHO may become hot enough to produce relativistic electrons, which changes the adiabatic index \pref{eq:gamma_def} from $\gamma = \gamma_\infty = 5/3$ to $\gamma = 13/9$.  
However, our derivation of \eref{eq:drho_dr} neglected gradients in $\gamma$, and we cannot use this equation to solve for $\rho(r)$ in the region where $\gamma(r)$ is varying.  
The solutions $\rho(r)$ and $T(r)$ from \eref{eq:outer} therefore do not apply to this inner adiabatic region, and we must derive new ones.  

%=========
At sufficiently small $r$ deep inside the inner adiabatic region where $T(r) \gtrsim O(\mathrm{few}) \times m_e$, the electrons in the accretion profile become relativistic and we have a constant $\gamma \approx 13/9$.  
Far outside the inner adiabatic region, on the other hand, we should have a constant $\gamma=5/3$.  
It can be seen from \eref{eq:outer} that for a constant adiabatic index $\gamma$ and in the region with $r > R$, the temperature profile $T(r)$ has a simple power law behavior of $1/r$, independent of $\gamma$.  
We take this as motivation for a matching procedure: since we know the behavior of $T(r)$ in the inner adiabatic regions where $T \gg m_e$ and the outer adiabatic where $T \ll m_e$, we construct a profile that approximately solves the Navier-Stokes equation for all $r$ by matching the two asymptotic solutions at where $T \approx m_e$.  
In particular, we perform the matching at $r=\Rrel$ where $T(\Rrel) = T_\mathrm{rel}$.  
We take $T_\mathrm{rel} = 2m_e/3$, because the ultra-relativistic and non-relativistic expressions for the plasma's total energy density are equal at this temperature~\cite{shapiro1973accretion}. 
We take $\rho_\mathrm{rel} = \rho^\mathrm{outer}(\Rrel)$ and $T_\mathrm{rel} = T^\mathrm{outer}(\Rrel)$ to ensure continuity of the density profile. 
Then the density and temperature profiles in the inner region are given by 
\begin{subequations}\label{eq:inner}
\begin{align}
%---
\label{eq:rho_inner}
	\rho(r) & = \rho_\mathrm{rel} \times \begin{cases}
	\left( \frac{3}{2}\frac{\Rrel}{R} - \frac{1}{2} \, \frac{r^2 \Rrel}{R^3} \right)^{\frac{1}{\gamma-1}} & , \ r \leq R < \Rrel ~,
	\vspace{2mm}\\
	\left( \frac{\Rrel}{r} \right)^{\frac{1}{\gamma-1}} & , \ R < r \leq \Rrel
	\end{cases} ~, \\ 
%---
\label{eq:T_inner}
	T(r) & = T_\mathrm{rel} \times \begin{cases}
	\frac{3}{2}\frac{\Rrel}{R} - \frac{1}{2} \, \frac{\Rrel r^2}{R^3}  & , \ r \leq R < \Rrel ~,
	\vspace{2mm}\\
	\frac{\Rrel}{r} & , \ R < r \leq \Rrel
	\end{cases} 
	\com
\end{align}
\end{subequations}
with $\gamma = 13/9$. We have also numerically solved the Navier-Stokes equations with a $r$-dependent $\gamma$ and found that the phenomenological solutions above agree well with the numerical solutions.  
Inspecting these solutions, we find that the density profile in the inner adiabatic region, $\propto r^{-9/4}$, is steeper than the density in the outer adiabatic region, $\propto r^{-3/2}$. 
This result follows from our assumption of hydrostatic accretion, and we discuss it further later, where we contrast with the Bondi accretion.  

%=========
\subsubsection{Outer isothermal region}
When we derived \eref{eq:drho_dr} from the more general fluid equations in \eref{eq:fluid_eqns}, we neglected the drag term $g_\mathrm{drag}$ and the cooling term $\dot{q}$.  
However for our study of dMACHOs in the early universe, the high density of the ambient plasma will make $\dot{q}$ important.  
(We will argue later that the drag term is still unimportant.)  
In this subsection we discuss the effect of a large $\dot{q}$.  

%=========
We return to the energy continuity equation \pref{eq:energy}, which is written as 
\begin{align} \label{eq:eos-coolling}
	\frac{v}{c_\infty} \, \rho^{\gamma_\infty-1} \, \RB \frac{d}{dr} \bigl( T / \rho^{\gamma_\infty-1} \bigr) = \Gamma \, \bigl( T_\mathrm{cmb} - T \bigr) 
	\qquad \text{where} \qquad 
	\Gamma \equiv \frac{8 \bar{x}_e \sigma_\mathrm{T} \rho_\mathrm{cmb}}{3 m_e (1 + \bar{x}_e)} \tB 
	\per
\end{align}
Here we've used \eref{eq:qdot_def} to write $\dot{q}$, used \eref{eq:gamma_def} to write $P = (\gamma-1)\Ecal$, used $\gamma = \gamma_\infty = 5/3$, used \eref{eq:E_from_rho} to write $\Ecal$ in terms of $T$, and multiplied both sides by $\tB$ from \eref{eq:RB_and_tB}.  
The term on the right-hand side represents a cooling of the accreted matter by scattering with relatively-cold CMB photons, and the dimensionless cooling factor $\Gamma$ evaluates to
\begin{align}
	\Gamma \simeq 2.74 
	\left( \frac{M}{1\Msun} \right) 
	\left( \frac{1+z}{1000} \right)^{5/2} 
	\left( \frac{T_\infty}{T_\mathrm{cmb}} \right)^{-3/2}
	\left( \frac{\bar{x}_e}{1} \right) 
	\left( \frac{1 + \bar{x}_e}{2} \right)^{-5/2} 
	\per
\end{align}
For $\Gamma \ll 1$, the cooling effects can be neglected, as we have assumed in the preceding sections.  
For $\Gamma \gg 1$, the cooling term is important at the Bondi scale where the derivative on the left-hand side of \eqref{eq:eos-coolling} evaluates to $d/dr \sim 1/\RB$, and here the cooling term enforces $T 
\approx T_\mathrm{cmb}$.  
The cooling term becomes less important at $r \ll \RB$ where the left-hand side grows relative to the right one like $v/r$.  
The tradeoff occurs where $r = r_\mathrm{cool}$ with $r_\mathrm{cool} \sim \Gamma^{-2/3} \RB$.  

%=========
In the isothermal region $T(r) = T_\mathrm{cmb} = T_\infty$ and the momentum continuity equation \pref{eq:momentum} gives 
\begin{align}
	\frac{\GN \widetilde{M}(r)}{r^2} + v(r) \frac{dv}{dr} + \frac{P_\infty}{\rho_\infty} \frac{d}{dr} \ln \rho(r) = 0  ~.
\end{align}
Note that the polytropic equation of state, given by \erefs{eq:P}{eq:T}, is inapplicable in the isothermal region.  
Solving the mass continuity equation \pref{eq:mass} gives $v(r) \propto 1/ [r^2 \rho(r)]$, and then the equation above can be solved for $\rho(r)$.  
A physically-reasonable, power-law solution is obtained for a maximal accretion rate~\cite{Ali-Haimoud:2016mbv}.  
The general solution can be written in terms of special functions, namely a product-log.  
The solution goes to a constant for $r > \RB$ and it behaves like a power law $\rho \propto r^{-3/2}$ for $ r < \RB$ where the logarithm term is negligible and before electrons become relativistic.  
This is the same power-law behavior that we encountered previously in the adiabatic approximation \pref{eq:rho_outer}.  

%=========
To summarize, we account for the isothermal region in the following way.  
If $\Gamma < 1$ then the cooling is negligible for all $r$, there is no isothermal region, and the profile in the outer adiabatic region are given by \eref{eq:outer}.  
If $\Gamma > 1$ then the cooling is important for $r > r_\mathrm{cool} = \Gamma^{-2/3} \RB$.  
We assume that cooling enforces $T(r) = T_\mathrm{cmb}$ for $r > r_\mathrm{cool}$, but meanwhile $\rho(r)$ is still given by \eref{eq:rho_outer}.  
Moving inward from the isothermal region to $r < r_\mathrm{cool}$, the density is still given by \eref{eq:rho_outer} and the temperature is given by \eref{eq:T} where the boundary condition $T(r_\mathrm{cool}) = T_\infty$ sets the value of $K$.  
In summary, we have 
\begin{subequations}\label{eq:cool}
\begin{align}
%---
\label{eq:rho_cool}
	\rho(r) & = \rho_\infty \times \begin{cases}
	\left(1 + \frac{3(\gamma_\infty-1)}{2}\frac{\RB}{R} - \frac{\gamma_\infty-1}{2} \, \frac{r^2 \RB}{R^3} \right)^{\frac{1}{\gamma_\infty-1}} & , \ r \leq R ~,
	\vspace{2mm}\\
	\left(1 + (\gamma_\infty-1) \frac{\RB}{r}\right)^{\frac{1}{\gamma_\infty-1}} & , \ R < r \lesssim \RB 
	\end{cases} ~, \\ 
%---
\label{eq:T_cool}
	T(r) & = T_\infty \times \begin{cases}
	\frac{1 + \frac{3(\gamma_\infty-1)}{2}\frac{\RB}{R} - \frac{\gamma_\infty-1}{2} \, \frac{r^2 \RB}{R^3}}{1 + (\gamma_\infty-1) \frac{\RB}{r_\mathrm{cool}}}  
	& , \ r \leq R ~
	\vspace{2mm}\\
	\frac{1 + (\gamma_\infty-1) \frac{\RB}{r}}{1 + (\gamma_\infty-1) \frac{\RB}{r_\mathrm{cool}}}  
	& , \ R < r \leq r_\mathrm{cool} ~
	\vspace{2mm}\\
	1 
	& , \ r_\mathrm{cool} < r ~
	\end{cases} \com
\end{align}
\end{subequations}
where $r_\mathrm{cool} = \Gamma^{-2/3} \RB$ and $\gamma_\infty = 5/3$.  
For the parameters of interest, we generally have $R < r_\mathrm{cool}$ as shown here.  

%=========
\subsubsection{Ionization of accreted matter}
Another assumption that went into \eref{eq:drho_dr} was a constant ionization fraction $x_e(r)$. 
Far from the dMACHO we have $\bar{x}_e = \lim_{r\to \infty} x_e(r)$, which can be $\bar{x}_e \ll 1$ if the ambient medium is mostly neutral or $\bar{x}_e \approx 1$ when the medium is already ionized.  
As one approaches the dMACHO, the temperature and density grow, and if the temperature rises to $O(\mathrm{eV})$ then the hydrogen atoms typically carry enough kinetic energy to ionize one another through scatterings such as $H + H \to H + e + p + \gamma$; this process is known as {\it collisional ionization}.  
Additionally, the liberated photons can carry enough energy to ionize other hydrogen atoms, $H + \gamma \to e + p + \gamma$, through a process known as {\it photoionization}.  
The temperature of the ionizing phase transition depends on environmental factors, such as the ionization fraction, but it is typically $T_\mathrm{ion} \sim \mathrm{eV}$ and we will take $T_\mathrm{ion} \simeq 1.5 \times 10^4 \kel \simeq 1.3 \eV$~\cite{Nobili:1991tx}.  
In general, ionization of the accreted matter is a complex, dynamical system with potentially significant backreaction on the density and temperature profiles.  
To quantify the uncertainty in our accretion calculation, we consider two cases in which ionization is accomplished entirely by \textit{either} collisional ionization \textit{or} photoionization~\cite{Ali-Haimoud:2016mbv}, where the former case leads to the more conservative limits and the latter more aggressive.  
In addition we assume that backreaction onto the density profile is negligible.  

\begin{itemize}
\item
If the accreted matter experiences only collisional ionization, it has been shown~\cite{shapiro1973accretion} that the temperature profile $T(r)$ experiences a plateau in the ionization region, and the density profile $\rho(r)$ is unaltered compared with the outer adiabatic region.  The ionization region corresponds to a range of radii $\Rend < r < \Rstart$.  The outer edge is defined by $T(\Rstart) = T_\mathrm{ion}$ where the temperature profile $T(r)$ is given by \eref{eq:T_outer}.  
The ionization profile $x_e(r)$ is calculated using the first law of thermodynamics, $\ud U = \ud Q$, which implies~\cite{Ali-Haimoud:2016mbv}
\begin{align}\label{eq:first_law}
	\frac{d}{dr} \left( \frac{3}{2} \bigl[ 1+x_e(r) \bigr] \, T(r) - \bigl[ 1-x_e(r) \bigr] \, E_\mathrm{I} \right) = -\bigl[ 1+x_e(r) \bigr] \, T(r) \, \rho(r) \, \frac{d\big(1/\rho(r)\big)}{dr} 
	\per
\end{align}
Here $E_\mathrm{I} \simeq 13.6 \eV$ denotes the binding energy of neutral hydrogen, $T(r) = T_\mathrm{ion}$, and $\rho(r)$ is given by \eref{eq:rho_outer}. 
Solving this equation gives
\begin{align}
\label{eq:xe_profile_CI}
	x_e(r)=(1+\bar{x}_e)\left(\frac{\rho(r)}{\rho(\Rstart)}\right)^{\left(\frac{3}{2}+\frac{E_I}{T_\mathrm{ion}}\right)^{-1}}-1\,,
\end{align}
where we have imposed the boundary condition $x_e(\Rstart) = \bar{x}_e$ to connect continuously onto the outer adiabatic region.  
The exponent is $(3/2+E_I / T_\mathrm{ion})^{-1} \simeq 0.0836 \approx 1/12$.  
The inner edge of the ionization region is defined by $x_e(\Rend) = 1$, or if this equation has no solution then effectively $\Rend = 0$.  
In summary, the ionization region extends from $\Rend < r < \Rstart$, and corresponds to a density profile $\rho(r)$ from \eref{eq:rho_outer}, a temperature profile $T(r) = T_\mathrm{ion}$, and an ionization fraction profile $x_e(r)$ from \eref{eq:xe_profile_CI}.  
Moving inward past the ionization region to $r_\mathrm{rel} < r < \Rend$, the density profile $\rho(r)$ is still given by \eref{eq:rho_outer}, the ionization fraction is $x_e(r) = 1$, and the temperature begins to rise again according to \eref{eq:T}, which implies $T(r) = T_\mathrm{ion} \, (\rho / \rho(\Rend))^{\gamma_\infty-1}$.  

\item
As a second ionization model, we suppose that the accreted matter is primarily ionized by photoionization.  
We assume that ionizing radiation produced at the inner accretion region is emitted isotropically and fully ionizes the surrounding accretion region out to a radius $r = \Rph$.  
We also assume that the ionization has a negligible backreaction on the temperature, so that the ionization fraction profiles can be approximated as 
\begin{align}
	\label{eq:xe_profile_PI}
	x_e(r) =\begin{cases}
	1 & , \ r < \Rph \vspace{1mm}\\
	\bar{x}_e & , \ \Rph < r 
	\end{cases} ~.
\end{align}
where $\rho(r)$ solves \eref{eq:drho_dr} as discussed in the last subsection.  For the inner region $r < R$, there is an approximately constant temperature with $T \approx (\RB/R) \, T_\infty$. The ionization threshold radius, $\Rph$, is determined by~\cite{osterbrock2006astrophysics} 
\begin{align}\label{eq:photo_ionization}
	\int_0^{\Rph} \! \ud r \ 4\pi r^2 \, n_e(r) \, n_p(r) \, \alpha_\mathrm{B}\big(T(r)\big)
	= \int_{E_\mathrm{I}/(2\pi)}^\infty \! \ud \nu \, \frac{L_\nu}{2\pi \nu} ~,
\end{align}
where $n_e(r) = n_p(r)$ are the electron and proton number densities from \eref{eq:ne_np}, $\alpha_\mathrm{B}(T)$ is the case-B recombination coefficient which characterizes the recombination rate \cite{osterbrock2006astrophysics}. We take $\alpha_\mathrm{B}(T) = (1.8\times 10^{-13}) (T_\mathrm{ion}/T(r))^{0.86}\,{\rm cm^{3}\,s^{-1}}$ where $T_\mathrm{ion} \simeq 1.3 \eV$~\cite{pequignot1991total}. $L_\nu$ is the luminosity spectrum (power per unit frequency) of the radiation.  
The calculation of $L_\nu$ will be discussed in the next subsection.
\end{itemize}

%=========
\subsubsection{Summary of results}
Based on the discussion above, it is possible to identify two special thresholds as the temperature increases from the outside of the profile to the inside:  the ionization temperature $T = T_\mathrm{ion}$ where ambient medium ionizes, and $T = T_\mathrm{rel}$ where the electrons transition from non-relativistic to relativistic.  
The configuration of the profile can thus be classified by whether the maximum temperature in the profile can reach these thresholds.  
Here we summarize the discussion above and show the expressions for $\rho(r)$, $T(r)$ and $x_e(r)$ for different cases.

%=========
The complete expressions of the $\rho$, $T$ and $x_e$ profiles have to be written as piecewise functions due to the temperature thresholds.  
For simplicity we assume $\RB$ is much larger than any other distance scale.  
The profile expressions in different cases of collisional ionization are given in Eqs.~(\ref{eq:CI_profile_2})-(\ref{eq:CI_profile_4}), in which $r_\mathrm{rel}$ is the radius where $T(r)=T_\mathrm{rel}$, the temperature when electrons become relativistic. 
$T_\mathrm{max}$ is the maximum temperature in the accretion profile, \textit{i.e.} $T_\mathrm{max}=T(r\rightarrow 0)$. 
Note that the temperature profile $T(r)$ in \eqref{eq:CI_profile_4} always follows the power-law $T\propto 1/r$ before and after $r_\mathrm{rel}$ and smoothly transit from one region to another, despite that it is defined in a piecewise way at $r=r_\mathrm{rel}$. We also note that if the CMB cooling effects are important with $\Gamma \gtrsim 1$, the profiles in the outer region with $r_\mathrm{cool} < r < R_\mathrm{B}$ should be replaced by  \eqref{eq:cool}.

%\newpage
%\begin{landscape}
\begin{subequations}
  \renewcommand{\arraystretch}{1.4}
  {\scriptsize
\begin{align}
%-----
%--
&
%\begin{table}
%\caption{caption}
\begin{array}{c c}
\hspace{-2mm}
\begin{array}{|c||c|c|}
	\hline
	\multicolumn{3}{|c|}{\text{\bf Collisional ionization}} \\ \hline
	\multicolumn{3}{|c|}{\text{\bf Case 1: }T_\mathrm{max}<T_\mathrm{ion} \simeq 1.3 \eV  }  \\ \hline
	r & [0, R) & [R, \infty) \\ \hline 
	\rho & \rho_\infty\left(\frac{\RB}{R} - \frac{1}{3} \, \frac{r^2 \RB}{R^3} \right)^{3/2} & \rho_\infty\left(1+\frac{2}{3} \frac{\RB}{r}\right)^{3/2} \\ \hline
	T & T_\infty\left(\frac{\RB}{R} - \frac{1}{3} \, \frac{r^2 \RB}{R^3} \right) & T_\infty\left(1+\frac{2}{3} \frac{\RB}{r}\right) \\ \hline
	x_e & \multicolumn{2}{c|}{\bar{x}_e} \\ 
	\hline
\end{array}  
%\label{eq:CI_profile_1} 
%\end{table} \\ 
%-----
%--
%&
\hspace{5mm}
\begin{array}{|c||c|c|c|}
\multicolumn{4}{c}{} \\
	\hline
	\multicolumn{4}{|c|}{\text{\bf Case 2: }T_\mathrm{max}=T_\mathrm{ion}} \\ \hline
	r & [0, R) & [R, \Rstart ) & [ \Rstart , \infty ) \\ \hline 
	\rho &  \rho_\infty\left(\frac{\RB}{R} - \frac{1}{3} \, \frac{r^2 \RB}{R^3} \right)^{3/2} & \multicolumn{2}{c|}{\rho_\infty\left(1+\frac{2}{3} \frac{\RB}{r}\right)^{3/2}} \\ \hline
		T & \multicolumn{2}{c|}{T_\mathrm{ion}} & T_\infty\left(1+\frac{2}{3} \frac{\RB}{r}\right) \\ \hline
	x_e & \multicolumn{2}{c|}{(1+\bar{x}_e)\left(\frac{\Rstart}{r}\right)^{0.125}-1} & \bar{x}_e \\ 
	\hline 
\end{array}\label{eq:CI_profile_2}
\end{array}
 \\ 
%-----
%--
&\begin{array}{|c||c|c|c|c|}
	\hline
	\multicolumn{5}{|c|}{\text{\bf Case 3: }T_\mathrm{ion}<T_\mathrm{max}<T_\mathrm{rel} = 2m_e/3 } \\ \hline
	r & [0, R) & [R, \Rend ) & [ \Rend, \Rstart ) & [ \Rstart , \infty ) \\ \hline 
	\rho & \rho_\infty\left(\frac{\RB}{R} - \frac{1}{3} \, \frac{r^2 \RB}{R^3} \right)^{3/2} & \multicolumn{3}{c|}{\rho_\infty\left(1+\frac{2}{3} \frac{\RB}{r}\right)^{3/2}} \\ \hline 
		T & T_\mathrm{ion}\left(\frac{3}{2}\frac{\Rend}{R}-\frac{1}{2}\frac{r^2 \Rend}{R^3}\right) & T_\mathrm{ion}\frac{\Rend}{r} & T_\mathrm{ion} & T_\infty\left(1+\frac{2}{3} \frac{\RB}{r}\right) \\ \hline
	x_e & \multicolumn{2}{c|}{1}  & (1+\bar{x}_e)\left(\frac{\Rstart}{r}\right)^{0.125}-1 & \bar{x}_e \\ 
	\hline
\end{array}\label{eq:CI_profile_3} \\ 
%-----
%--
&\begin{array}{|c||c|c|c|c|c|}
	\hline
	\multicolumn{6}{|c|}{\text{\bf Case 4: }T_\mathrm{max}>T_\mathrm{rel} } \\ \hline
	r & [0, R) & [R, r_\mathrm{rel}) & [r_\mathrm{rel}, \Rend) & [\Rend, \Rstart) & [\Rstart, \infty) \\ \hline 
	\rho & \rho_\infty\left(\frac{2}{3} \frac{\RB}{r_\mathrm{rel}}\right)^{3/2}\left(\frac{3}{2}\frac{r_\mathrm{rel}}{R}-\frac{1}{2}\frac{r^2r_\mathrm{rel}}{R^3}\right)^{9/4} & \rho_\infty\left(\frac{2}{3} \frac{\RB}{r_\mathrm{rel}}\right)^{3/2}\left(\frac{r_\mathrm{rel}}{r}\right)^{9/4} &\multicolumn{3}{c|}{\rho_\infty\left(1+\frac{2}{3} \frac{\RB}{r}\right)^{3/2}} \\ \hline
	T & T_\mathrm{rel} \left(\frac{3}{2}\frac{r_\mathrm{rel}}{R}-\frac{1}{2}\frac{r^2r_\mathrm{rel}}{R^3}\right) & T_\mathrm{rel} \frac{r_\mathrm{rel}}{r} & T_\mathrm{ion}\frac{\Rend}{r} & T_\mathrm{ion} & T_\infty\left(1+\frac{2}{3} \frac{\RB}{r}\right) \\ \hline
	x_e & \multicolumn{3}{c|}{1} & (1+\bar{x}_e)\left(\frac{\Rstart}{r}\right)^{0.125}-1 & \bar{x}_e \\ 
	\hline
\end{array}\label{eq:CI_profile_4} 
\end{align}
}
\end{subequations}

%\end{landscape}

%=========
The situation is simpler for photoionization as there are only two cases, and the profile functions are shown in \erefs{tab:PI_1}{tab:PI_2}.  
Since there is no ionization plateau in the case of photoionization, the core temperature $T_\mathrm{max}$ may be larger than for collisional ionization, leading to stronger signals and/or constraints.  
\begin{subequations}
  \renewcommand{\arraystretch}{1.4}
{\scriptsize
\hspace{-15mm}
\begin{align}
%-----
%--
&\begin{array}{|c||c|c|c|}
	\hline
	\multicolumn{4}{|c|}{\text{ \bf Photoionization}} \\ \hline
	\multicolumn{4}{|c|}{\text{\bf Case 1: }T_\mathrm{max}<T_\mathrm{rel} =2 m_e /3} \\ \hline
	r & [0, R) & [R, \Rph) & [\Rph, \infty) \\ \hline 
	\rho & \rho_\infty\left(\frac{\RB}{R} - \frac{1}{3} \, \frac{r^2 \RB}{R^3} \right)^{3/2} & \multicolumn{2}{c|}{\rho_\infty\left(1+\frac{2}{3} \frac{\RB}{r}\right)^{3/2}} \\ \hline
	T & T_\infty\left(\frac{\RB}{R} - \frac{1}{3} \, \frac{r^2 \RB}{R^3} \right) & \multicolumn{2}{c|}{T_\infty\left(1+\frac{2}{3} \frac{\RB}{r}\right)} \\ \hline
	x_e & \multicolumn{2}{c|}{1} & \bar{x}_e \\ \hline
\end{array}\label{tab:PI_1} \\ 
%--
&\begin{array}{|c||c|c|c|c|}
	\hline
	\multicolumn{5}{|c|}{\text{\bf Case 2: }T_\mathrm{max}>T_\mathrm{rel}  } \\ \hline
	r & [0, R) & [R, r_\mathrm{rel}) & [r_\mathrm{rel}, \Rph) & [\Rph, \infty) \\ \hline 
	\rho & \rho_\infty\left(\frac{2}{3} \frac{\RB}{r_\mathrm{rel}}\right)^{3/2}\left(\frac{3}{2}\frac{r_\mathrm{rel}}{R}-\frac{1}{2}\frac{r^2r_\mathrm{rel}}{R^3}\right)^{9/4} & \rho_\infty\left(\frac{2}{3} \frac{\RB}{r_\mathrm{rel}}\right)^{3/2}\left(\frac{r_\mathrm{rel}}{r}\right)^{9/4} & \multicolumn{2}{c|}{\rho_\infty\left(1+\frac{2}{3} \frac{\RB}{r}\right)^{3/2}} \\ \hline
	T & \frac{2}{3}m_e \left(\frac{3}{2}\frac{r_\mathrm{rel}}{R}-\frac{1}{2}\frac{r^2r_\mathrm{rel}}{R^3}\right) & \frac{2}{3}m_e \frac{r_\mathrm{rel}}{r} & \multicolumn{2}{c|}{T_\infty\left(1+\frac{2}{3} \frac{\RB}{r}\right)} \\ \hline
	x_e & \multicolumn{3}{c|}{1} & \bar{x}_e \\ \hline
\end{array} \label{tab:PI_2}
\end{align}
}
\end{subequations}

\begin{figure}[p] %[tbp]
%	\centering
\begin{center}
	\includegraphics[width=0.47\linewidth]{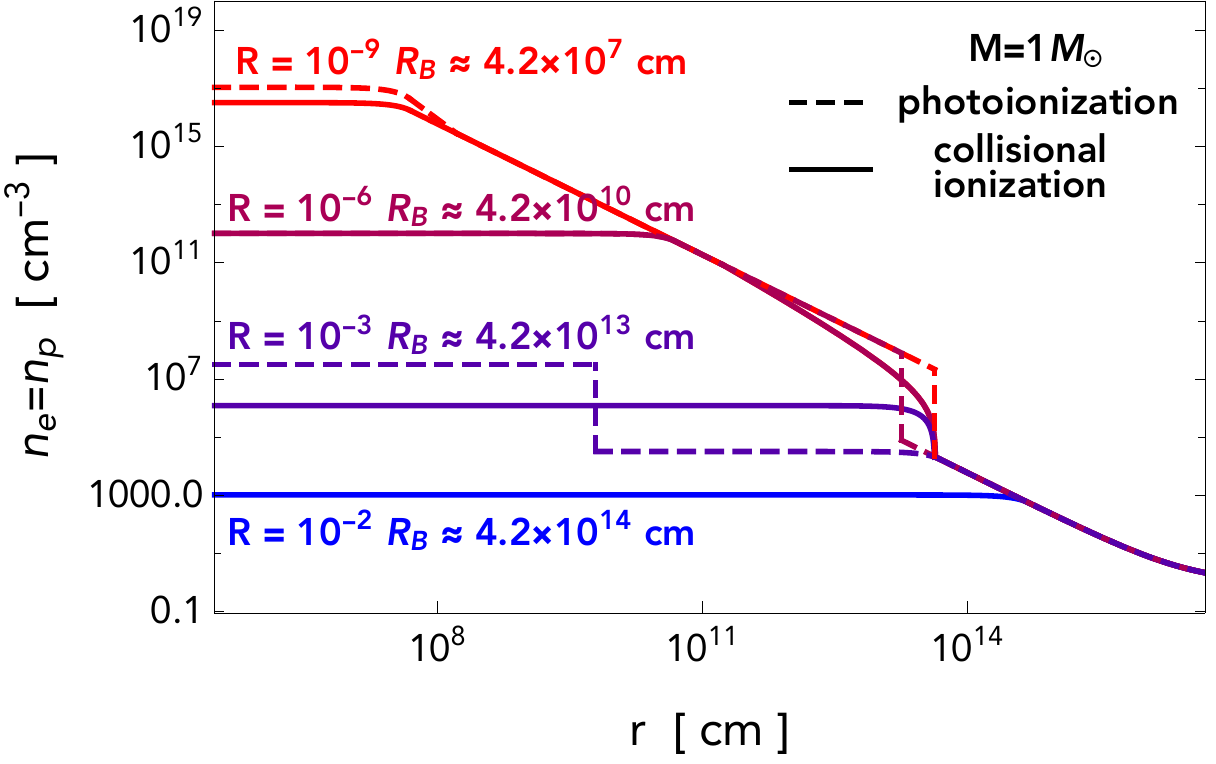} 
\quad \hspace{3mm}
	\includegraphics[width=0.47\linewidth]{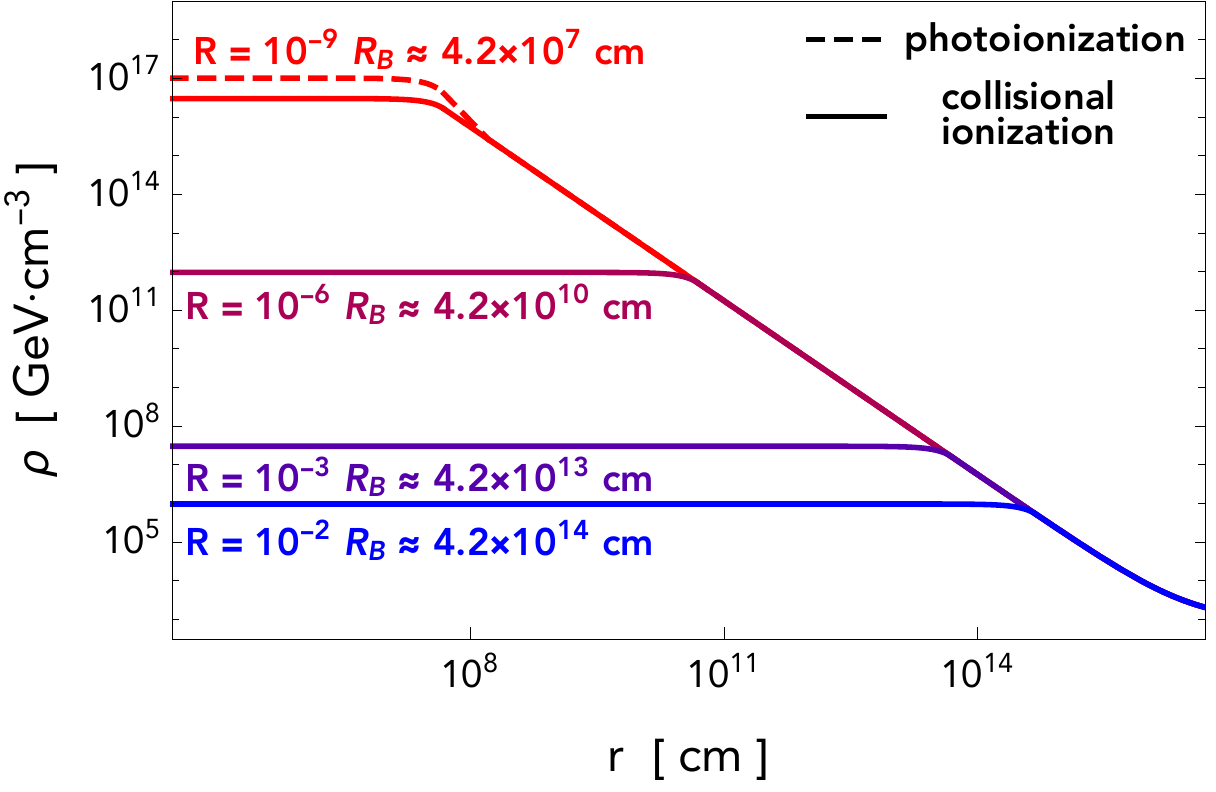} \\ \vspace{0.5cm}
	\includegraphics[width=0.47\linewidth]{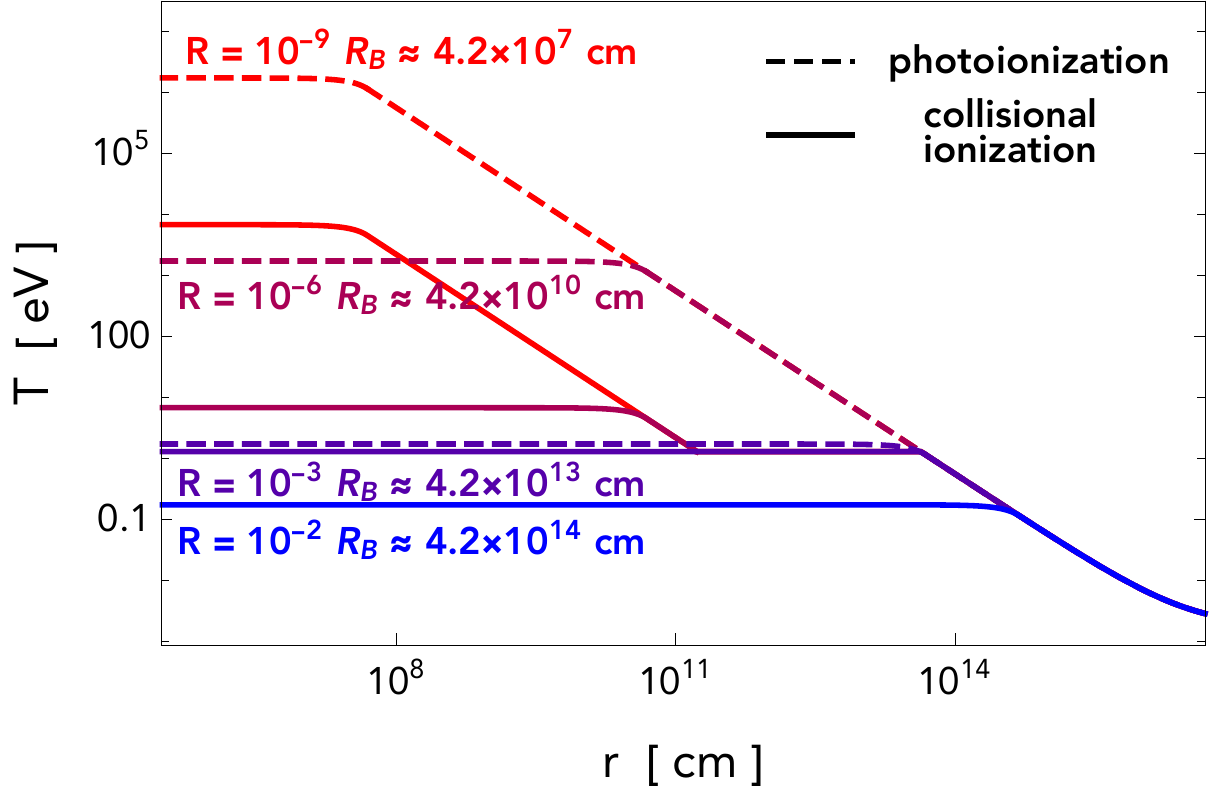} 
\quad \hspace{3mm}
	\includegraphics[width=0.47\linewidth]{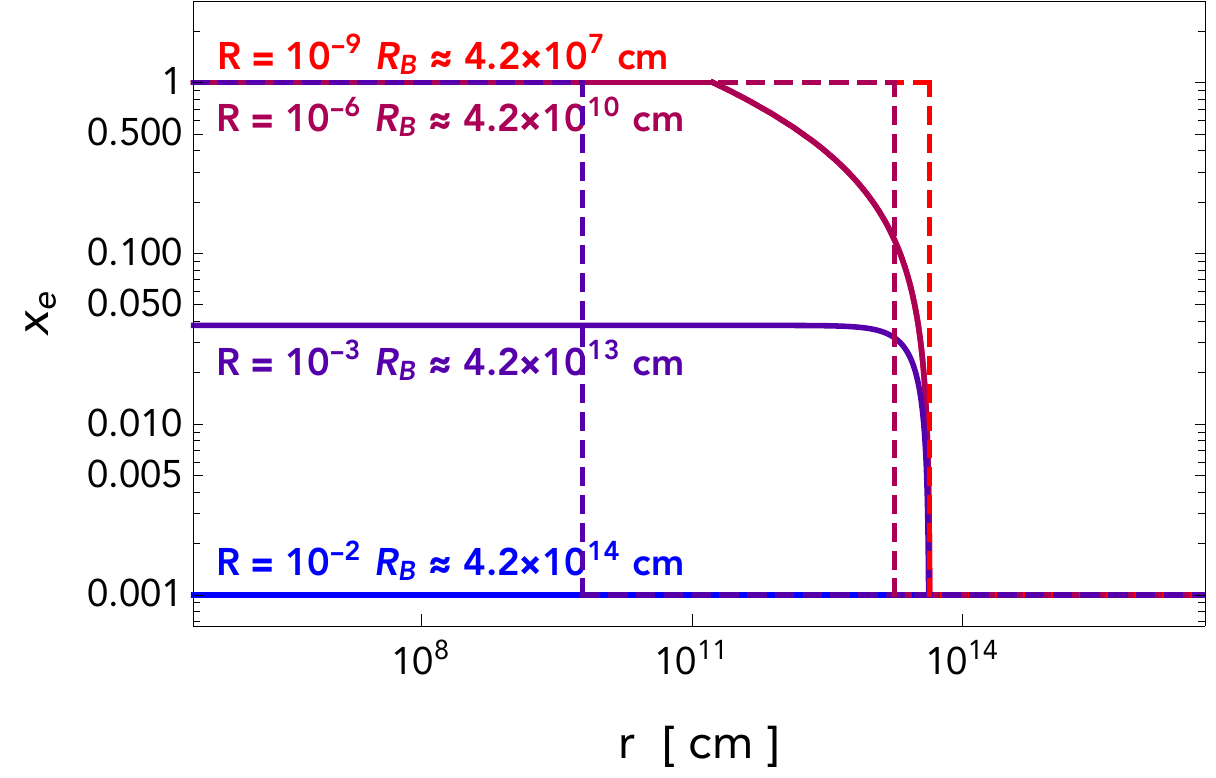} 
	\caption{\label{fig:profiles}
	The electron and proton number densities $n_e(r) = n_p(r)$, mass density $\rho(r)$, temperature $T(r)$, and ionization fraction $x_e(r)$ as a function of the distance $r$ away from the center of a spherically-symmetric dMACHO. For the illustration purpose, the relative velocity between dMACHOs and the ambient gas is ignored here, but will be kept for our later signal calculations. The four sets of colored curves correspond to different dMACHO models with mass $M = 1 \Msun$ and variable radius $R$.  Different models for the ionization of accreted matter are shown: collisional ionization (solid) and photoionization (dashed). 
Here, $\rho_\infty = 938\,{\rm GeV\,cm^{-3}}$, $T_\infty = 2 \times 10^{-3} \eV$, and $\bar{x}_e = 10^{-3}$, which implies $\RB = 4.2\times 10^{16} \cm$ from \eref{eq:RB_and_tB}. Also we assume $g_\mathrm{drag} = 0$, and $\dot{q} = 0$. Observational probes of dMACHOs will be most sensitive to compact dMACHOs with high-temperature, ionized cores.  
	}
\end{center}
\end{figure}

%=========
The results of this subsection are also summarized in \fref{fig:profiles}.  
We show profile functions for the density $n_e(r) = n_p(r)$, temperature $T(r)$, and ionization fraction $x_e(r)$ for several representative dMACHO models.  
The models are defined by the uniform-density mass profile in \eref{eq:hard_sphere} with mass $M = 1 \Msun$ and radius $R$.  
We have taken $\rho_\infty = 938\,{\rm GeV\,cm^{-3}}$, $T_\infty = 2 \times 10^{-3} \eV$, and $\bar{x}_e = 10^{-3}$, which implies $K = 2.2\times 10^{-14}\,{\rm GeV^{-2/3}\,cm^2}$ from \eref{eq:T}, $P_\infty = 2\times 10^{-9}\,{\rm GeV\,cm^{-3}}$ from \eref{eq:P_from_rho}, $c_\infty = 1.9\times 10^{-6}$ from \eref{eq:c_infty}, and $\RB = 4.2\times 10^{16} \cm$ from \eref{eq:RB_and_tB}. The boundary condition is chosen to match the environment of a molecular cloud discussed later in this section.
Examining the figure, we see that increasing the dMACHO density, by decreasing $R$ at fixed $M$, leads to more accretion, a higher temperature $T(r=0)$, and greater ionization $x_e(r=0)$.  
A relatively extended dMACHO (large $R$) accretes less efficiently, and may not even reach $T(r) = T_\mathrm{ion}$ where collisional ionization can occur, whereas a relatively compact dMACHO (small $R$) can fully ionize the baryonic matter in its vicinity.  

Here and in the following sections we will focus on only the uniform-density dMACHO mass profile from \eref{eq:hard_sphere}.  We have also checked that the profiles of accreted baryonic matter are qualitatively unchanged for different dMACHO mass profiles.  As a result, we expect the accretion-based dMACHO constrained, derived below, to be similar for other mass profiles. 

%=========

%=========
\subsubsection{Comparing accretion onto dMACHOs and black holes}

%=========
At this point it is worthwhile to compare our calculation here, for accretion onto a dMACHO, with similar calculations in the literature for accretion onto black holes.  
The mass continuity equation in \eref{eq:mass} is solved by a stationary accretion flow with speed $v(r) = - \dot{M} / [4 \pi r^2 \rho(r)]$ where $\dot{M}$ is called the mass accretion rate.  
It is customary to write $\dot{M} = \lambda \times 4 \pi \RB^2 \rho_\infty c_\infty$, which expresses the accretion rate in terms of the dimensionless variable $\lambda$.  
Accretion onto a black hole is usually modeled with Bondi accretion~\cite{Bondi:1952ni}, corresponding to a maximal accretion rate; \textit{e.g.} $\lambda = 1/4$ for $\gamma = 5/3$ and negligible $g_\mathrm{drag}$ and $\dot{q}$.  
It is reasonable to apply Bondi accretion to the study of black holes, because the infalling matter can be absorbed by the black hole at its horizon, at least for an astrophysical-scale black hole with a tiny Hawking temperature.  
However, we would argue that Bondi accretion is not the appropriate model for accretion onto a dMACHO.  
Since there is no event horizon, the infalling matter is not absorbed, but rather it must bounce back or flow outward, implying a time-dependent solution.  
Instead we have used the hydrostatic approximation to study accretion onto dMACHOs in this work.  
In terms of the dimensionless accretion rate, the hydrostatic approximation corresponds to the limit $\lambda \to 0$.  
Intuitively, the build up of accreted matter around the dMACHO provides a radiation pressure that supports a static configuration with negligible flow velocity.  

%=========
\begin{figure}[tbp]
\begin{center}
	\includegraphics[width=0.6\linewidth]{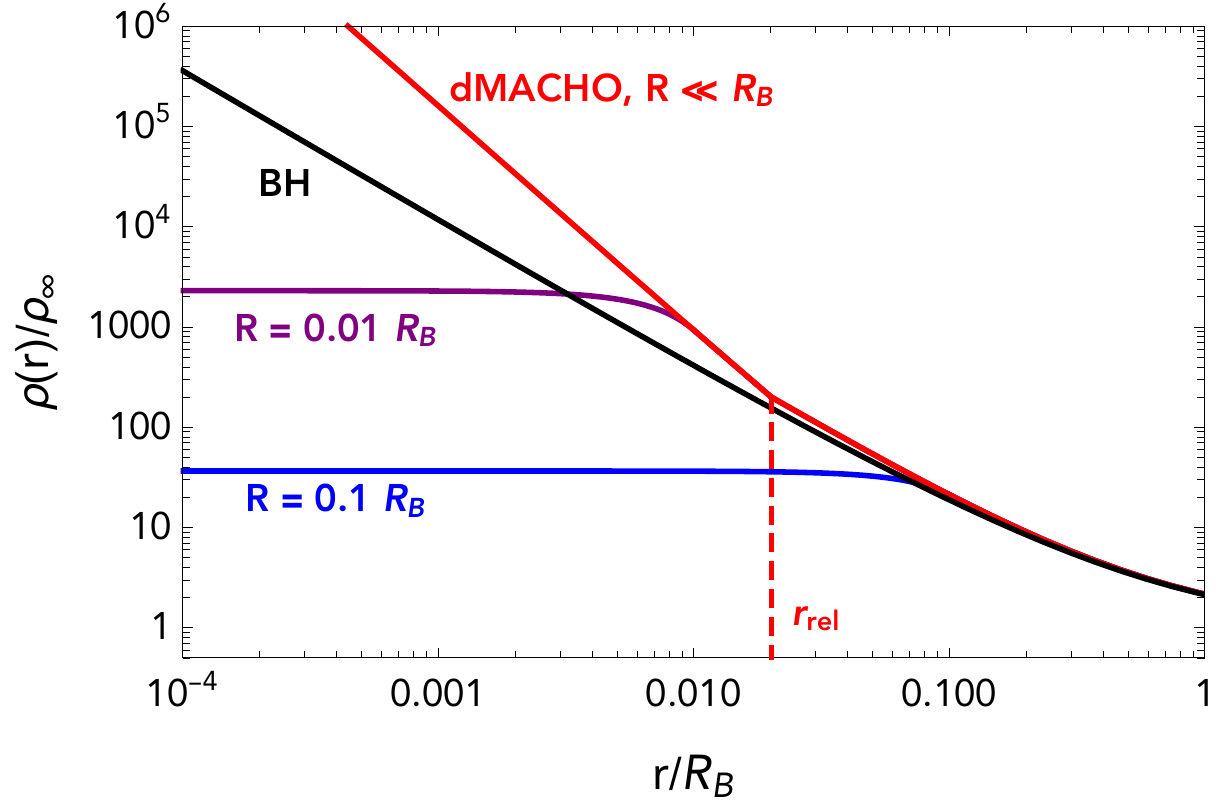} 
	\caption{\label{fig:Hydro_BH_cmp}
We show mass density profiles of baryonic matter that has accreted onto a dMACHO and a black hole (BH).  The dMACHO profiles are calculated using the hydrostatic approximation, and the BH profiles are calculated using the Bondi approximation.  
The profiles are calculated including both the outer adiabatic region ($\gamma=5/3$) and the inner adiabatic region ($\gamma=13/9$), and assuming $g_\mathrm{drag} = 0$, and $\dot{q} = 0$. The colored curves correspond to a dMACHO with radius $R = \RS \ll \RB$ (red), $R = 0.01 \, \RB$ (purple), and $R = 0.1 \, \RB$ (blue), while the black curve corresponds to a BH.  
Note that $\rho \propto r^{-3/2}$ for the BH and the outer adiabatic region of the ultra-compact dMACHO, but $\rho \propto r^{-9/4}$ for the inner adiabatic region of the ultra-compact dMACHO. The transition radius, $r_{\rm rel}$, is when electrons change from non-relativistic (outer) to relativistic (inner) ones. Parameters are chosen for sake of illustration; in practice $R \ll \RB$.
}
\end{center}
\end{figure}
 
%=========
To understand quantitatively how the two accretion scenarios differ, we have calculated the density profiles for both hydrostatic and Bondi accretion, and we present these results in \fref{fig:Hydro_BH_cmp} with $g_\mathrm{drag} = 0$, and $\dot{q} = 0$.  
In the outer adiabatic region where $\gamma = 5/3$, the Bondi solution gives $\rho(r) \propto r^{-3/2}$ for $r \ll \RB$~\cite{Bondi:1952ni}, and we have already seen in \eref{eq:rho_outer} and \fref{fig:profiles} that the hydrostatic solution gives $\rho(r) \propto r^{-3/2}$ for $r_\mathrm{rel} < r  \ll \RB$.
The dMACHO's hydrostatic density profile approaches the black hole's Bondi density profile, but remains larger by a constant factor of about $35\%$.  
This can be understood from the Navier-Stokes momentum equation \pref{eq:momentum} where in the case of Bondi accretion, the convective acceleration term, $\rho \, v \, v^\prime$, compensates some of the gravitational pressure, leading to less efficient accretion.  

%=========
On the other hand, if the temperature of the accreting matter exceeds $T = T_\mathrm{rel} = 2m_e/3$ then there is a more pronounced difference between the dMACHO and BH accretion profiles.  One can see this difference for the region with $ r < r_\mathrm{rel}$ where the inner adiabatic region has $\gamma = 13/9$ instead of $5/3$.   
For Bondi accretion, the density profile has the same scaling here as in the outer adiabatic region, namely $\rho \propto r^{-3/2}$, and the temperature profile is given by \eref{eq:T} with $\gamma = 13/9$.  
However for hydrostatic accretion, which we use to model the dMACHO, the density profile steepens to $\rho \propto r^{-9/4}$.  
One can understand this result intuitively, because the hydrostatic solution needs a larger density at $r \approx 0$ to provide the pressure that supports the surrounding matter from collapsing inward.  
This broken-power-law behavior distinguishes our analysis from previous one, for instance, Ref.~\cite{Savastano:2019zpr}.  
Due to the increased density and temperature, the signatures of accretion are expected to be stronger for dMACHOs than for BHs.  
Other applications of hydrostatic accretion onto dark matter have been studied in Refs.~\cite{Bai:2016wpg,Curtin:2019ngc}.

%==================================
% Radiation from the accreted matter
%==================================
\subsection{Radiation from the accreted matter}\label{sec:emission}

%=========
Matter that is accreted onto a dMACHO is heated up and begins to radiate.  
In this subsection we present the formulas that are used to calculate the spectrum of that radiation, and in the following subsections we discuss how the radiation can be used to test dMACHOs.  

%=========
\subsubsection{Spectrum and luminosity of radiation}
Consider an isolated, spherical dMACHO that accretes from the surrounding electron-proton plasma, as described in the previous subsection.  
The accreted electrons and protons may scatter, producing an associated bremsstrahlung radiation via processes like $ee \rightarrow ee \gamma$ and $ep \rightarrow ep\gamma$.  
The emissivity of this radiation $j_\nu(r)$ (emission power per volume per frequency per steradian) into photons of frequency $\nu$ at a distance $r$ away from the dMACHO is given by~\cite{draine2010physics} 
\begin{align}
\label{eq:emissivity}
	j_\nu(r) = \frac{8}{3} \biggl( \frac{2\pi \, m_e}{3 \, T(r)} \biggr)^{1/2} \frac{\alpha^3}{m_e^2} \, g_{ff}\bigl(\nu,\, T(r)\bigr) \, e^{-2\pi\nu/T(r)} \, n_e(r) \, n_p(r) 
	\com
\end{align}
where $\alpha \simeq 1/137$ is the electromagnetic fine structure constant.
Recall that $n_e(r)$ and $n_p(r)$ are the electron and proton number density profiles from \eref{eq:ne_np} and $T(r)$ is the temperature profile from \eref{eq:T}.  
The dimensionless factor, $g_{ff}(\nu,T)$, which is known as the free-free Gaunt factor~\cite{deAvillez:2015kra}, accounts for quantum corrections.
For the accretion profile discussed above, $g_{ff}(\nu,T)$ mainly comes from electron-electron ($e$-$e$) and electron-proton ($e$-$p$) scattering.
Ref.~\cite{nozawa2009analytic} provides sub-percent accuracy level fitting formulae for the contribution from $e$-$e$ scattering, while in Ref.~\cite{quigg1968electron} analytic expressions for the contribution from $e$-$p$ scattering in the non-relativistic and extreme-relativistic limits are given. We adopt the sum of the two contributions to $g_{ff}(\nu,T)$ for our later calculation.~\footnote{For the $e$-$p$ scattering, when $T/m_e<0.3$ we use the non-relativistic result in \cite{nozawa2009analytic}, and when $T/m_e>0.3$ we use the extreme-relativistic expression therein.}
Also, it is convenient to define the thermally averaged Gaunt factor by 
\begin{align}\label{eq:gff_approx}
	\langle g_{ff}(T) \rangle \equiv \frac{2\pi}{T} \int_0^\infty \! \! \ud \nu \ g_{ff}(\nu,T) \ e^{-2 \pi \nu / T}
	\com
\end{align}
which has a percent-accuracy level fitting formula given in Ref.~\cite{Ali-Haimoud:2016mbv}.
Using \eref{eq:gff_approx} we can evaluate the frequency integral of the emissivity to obtain the radiation power density (radiation power per volume), which is 
\begin{align}\label{eq:L_total}
	\mathcal{L}(r) 
	& = \int \! \ud \Omega \int_0^\infty \! \ud \nu \, j_\nu(r) 
	= \frac{16}{3} \biggl( \frac{2 \pi m_e\,T(r)}{3} \biggr)^{1/2} \frac{\alpha^3}{m_e^2} \, \big\langle g_{ff} (T(r))\big\rangle \, n_e(r) \, n_p(r)
	\com
\end{align}
where we have also integrated the isotropic emission over solid angle, which brings a factor of $\int \! \ud \Omega = 4\pi$.  
The luminosity spectrum $L_\nu$ (power per frequency) is obtained by integrating the emissivity over space 
\begin{align}\label{eq:L_nu}
	L_\nu = \int \! \ud \Omega \, \int_0^\infty \! \ud r \, 4 \pi r^2 \, j_\nu(r) 
	\com
\end{align}
where $\int \! \ud \Omega = 4\pi$, and the total luminosity is then
\begin{align}\label{eq:L}
	L = \int_0^\infty \! \ud \nu \, L_\nu = \int_0^\infty \! \ud r \, 4\pi r^2 \, \mathcal{L}(r)
	\per   
\end{align}
The temperature and density profiles must first be calculated before these integrals can be performed.  

With the expression given in \eref{eq:CI_profile_2}-\eqref{eq:CI_profile_4}, \eqref{tab:PI_1} and \eqref{tab:PI_2}, it is possible to derive some analytical expression for \eref{eq:L}. For example, when approximating $\big\langle g_{ff} (T(r))\big\rangle\approx 1$, and assuming $\bar{x}_e=1$, the luminosity $L$ in the case \eref{eq:CI_profile_3} (or equivalently \eqref{tab:PI_1} as there is not an ionization region) can be well approximated as 
\begin{align}\label{eq:lum_analytic}
L\approx 2.4\times 10^{-5}\times\frac{\rho^2_\infty T^{1/2}_\infty \RB^{7/2}}{R^{1/2}\,m^2_p\,m^{3/2}_e } = 4.1\times 10^{-6}\,\times\,\frac{\rho^2_\infty\,m^{3/2}_p\,(G_NM)^{7/2}}{T_\infty^{3}\,m^{3/2}_e \,R^{1/2}}  \,.
\end{align}

%=========
Note that the emissivity in \eref{eq:emissivity} is nonzero for a homogeneous electron-proton plasma, even in the absence of accretion.  
Of course this contribution to $j_\nu$ simply captures the emission of radiation that keeps the plasma in thermal equilibrium at temperature $T$.  
To determine the enhanced emission that arises from accretion onto the dMACHO, the quantities of interest are instead 
\begin{align}
%---
	\label{eq:L_total_prime}
	\mathcal{L}(r) & = \int \! \ud \Omega \, \int_0^\infty \! \ud \nu \, \Bigl[ j_\nu(r) - j_\nu \bigr|_\infty \Bigr] 
	\tag{\ref*{eq:L_total}$^\prime$} ~, \\ 
%---
	\label{eq:L_nu_prime}
	L_\nu & = \int \! \ud \Omega \, \int_0^\infty \! \ud r \, 4 \pi r^2 \, \Bigl[ j_\nu(r) - j_\nu \bigr|_\infty \Bigr] 
	\tag{\ref*{eq:L_nu}$^\prime$} ~, \\ 
%---
	\label{eq:L_prime}
	L & = \int \! \ud \Omega \, \int_0^\infty \! \ud r \, 4 \pi r^2 \, \int_0^\infty \! \ud \nu \, \Bigl[ j_\nu(r) - j_\nu \bigr|_\infty \Bigr] 
	\tag{\ref*{eq:L}$^\prime$}  ~,
\end{align}
where $\int \! \ud \Omega = 4\pi$, and $j_\nu(r)$ is given by \eref{eq:emissivity} and $j_\nu|_\infty = \lim_{r \to \infty} j_\nu(r)$.  
If the dMACHO were not present, we would find $\mathcal{L}(r) = L_\nu = L = 0$.  
Note that our assumptions of stationary accretion, discussed in the previous subsection, are only reliable for $r \lesssim \RB$, and so we cutoff the $\ud r$ integrals at $r = \RB$ in practice; the integral is typically dominated by $r \sim R \ll \RB$.  
We have checked that the accreted matter is not optically thick, \textit{i.e.} the optical depth is $\tau = \int_0^\infty \ud r \, n_e \, \sigma_\mathrm{T} \ll 1$, and we expect that most of the radiation does not re-scatter.  

%=========
To calculate CMB observables in the next section, we assume a uniform population of dMACHOs that all have a common mass and radius, $M$ and $R$.  
The luminosity from a given dMACHO at redshift $z$ is written as $L(z)$ and calculated using \eref{eq:L_prime} where the redshift dependence enters through the boundary conditions in \eref{eq:BCs}.  
Then the power density at redshift $z$ is written as 
\begin{align}\label{eq:P_def}
	P(z) = L(z) \, n_\mathrm{dMACHO}(z)  ~,
\end{align}
where $n_\mathrm{dMACHO}(z)$ is the population number density of dMACHOs at redshift $z$.  
Assuming that dMACHOs make up all of the dark matter, we can write $n_\mathrm{dMACHO}(z) = (\Omega_\DM h^2) \, (3 \Mpl^2 H_{100}^2) \, (1+z)^3 / M$ where $\Omega_\DM h^2 \simeq 0.12$ and $H_{100} \equiv 100 \km / \mathrm{sec} / \mathrm{Mpc}$.  

%=========
\subsubsection{Relative velocity between dMACHOs and the thermal bath}
In our previous study of accretion onto dMACHOs, we have implicitly assumed that dMACHOs are at rest with respect to the ambient medium.  
More precisely, we have assumed that the relative speed between the dMACHO and the medium, $v_\mathrm{rel}$, is small compared to the adiabatic sound speed of the medium, $c_\infty$ from \eref{eq:c_infty}.  
However this is not always the case for the systems of interest.  
A larger $v_\mathrm{rel}$ makes it harder for the accreting matter to be captured in the gravitation potential of the dMACHO.  
This decreases the size of the accretion region and the luminosity of the accreted matter.  

%=========
To account for a finite $v_\mathrm{rel}$, we follow the approach that was suggested by \rref{Bondi:1952ni}.  
Namely, we replace $c_\infty \to \sqrt{c_\infty^2 + v_\mathrm{rel}^2}$ when evaluating the Bondi radius with \eref{eq:RB_and_tB}.  
Since different dMACHOs will move with different speeds and in different environments, the value of $v_\mathrm{rel}$ is treated as a stochastic variable whose probability distribution depends on the system under consideration.  
Following \rref{Ali-Haimoud:2016mbv} we assume that on large scales the relative velocity, $\vec{v}_\mathrm{rel}$, follows a three-dimensional Gaussian linear distribution with standard deviation $\langle v_\mathrm{L}^2 \rangle^{1/2}$.  
For dMACHOs on cosmological scales at redshift $z$, we have~\cite{Tseliakhovich:2010bj} 
\begin{align}\label{eq:vL}
	\langle v_\mathrm{L}^2 \rangle^{1/2} = \mathrm{min}\bigl[1,z/10^3\bigr]\times 30 \km / \mathrm{sec} ~,
\end{align}
whereas for dMACHOs in virialized galactic halos today we expect a value closer to $\langle v_\mathrm{L}^2 \rangle^{1/2} \approx 300 \km / \mathrm{sec}$.  

%=========
For the calculation of observables, it is necessary to marginalize over the distribution of $v_\mathrm{rel}$ values.  
For an observable such as the luminosity, $\Ocal = L$ from \eref{eq:L_prime}, or the spectrum, $\Ocal = L_\nu$ from \eref{eq:L_nu_prime}, or the power density, $\Ocal = P$ from \eref{eq:P_def}, we evaluate the velocity-averaged observable as 
\begin{align}\label{eq:L_ave}
	\langle \Ocal \rangle = \frac{4\pi}{(2\pi\langle v^2_\mathrm{L}\rangle/3)^{3/2}} \int^\infty_0 \! \ud v_\mathrm{rel} \, v_\mathrm{rel}^2 \, e^{-\frac{v_\mathrm{rel}^2}{2\langle v_\mathrm{L}^2 \rangle/3}} \ \Ocal \bigr|_{c_\infty \to \sqrt{ c_\infty^2 + v_\mathrm{rel}^2 }} ~.
\end{align}
In Sec.~\ref{sec:CMB} we drop the angled brackets to simplify notation, but all of our calculations use this averaging.  

%=========
\subsubsection{Self-consistency check for luminosity of radiation}
For our solution to be self-consistent, the energy liberated from the dMACHO as radiation must not exceed the energy provided to the dMACHO as accreted matter.  
As matter falls from the outer edge of the accretion region at $r = \RB$ down to the dMACHO's surface at $r = R$, the matter's gravitational potential energy is converted into kinetic energy.  
Requiring the total luminosity $L$ to be smaller than the rate of kinetic energy deposition leads to 
\begin{align}
	L \leq \left( \frac{\GN M}{R} - \frac{\GN M}{\RB} \right) \, \bigl( 4 \pi \RB^2 \bigr) \, \rho(\RB) \, \big| v(\RB) \big|
	\per
\end{align}
To derive this formula we have allowed for a nonzero flow velocity $\vec{v}(r) = v(r) \, \vec{r}/r$, and we have used the mass continuity equation \eref{eq:mass} to write $4 \pi R^2 \rho(R) v(R) = 4 \pi \RB^2 \rho(\RB) v(\RB)$.  
The flow speed is bounded from above by the asymptotic sound speed, $v(\RB) < c_\infty$. 
In our later calculations, we use $v(\RB) = c_\infty$ for estimation, and find our calculated luminosity ({\it e.g.} left panel of \fref{fig:Dxe}) to be much smaller than the kinetic energy deposition. 
Moreover, the average over relative velocity $v_{\rm rel}$ in Eq.~\eref{eq:L_ave} is still consistent with the hydrostatic approximation, as the typical value of $v_{\rm rel}$ defined in Eq.~\eref{eq:vL} is comparable to the sound speed at infinity $c_\infty$ (see Fig.7 of Ref.~\cite{Ali-Haimoud:2016mbv} for a comparison between $\langle v_\mathrm{L}^2 \rangle^{1/2}$ and $c_\infty$).
%\AL{Add some text to say that $v_{rel}$ doesn't invalidate the hydrostatic approx, since anyway $v_{rel} \sim c_\infty$.}

%==================================
% Effects on the cosmic microwave background
%==================================
\subsection{Effects on the cosmic microwave background}\label{sec:CMB}

%=========
The accretion of baryonic matter onto dMACHOs in the early universe may leave an imprint on the cosmic microwave background radiation through its effect on the CMB's spectrum and pattern of anisotropies.  
In this section we assess the ability of CMB measurements to test dMACHOs.  

%=========
\subsubsection{Spectral distortions}  
The hot accreting matter provides a source of energy injection into the primordial plasma.  
Such energy injections can lead to distortions in the spectrum of the cosmic microwave background radiation~\cite{Sunyaev:1970er}.  
Spectral distortions arising from primordial black holes have been studied previously by Refs.~\cite{Ricotti:2007au,Blum:2016cjs,Ali-Haimoud:2016mbv}.  
They found that spectral distortions at the level probed by COBE-FIRAS~\cite{Fixsen:1996nj} do not constrain PBH dark matter for $M \lesssim 10^4 \Msun$, due to the strong dragging and cooling from the CMB in the early universe. For the same reason, CMB spectral distortions are not expected to impose constrains on dMACHO parameter space with $M \lesssim 10^4 \Msun$. 

%=========
\subsubsection{Anisotropies}  
\begin{figure}[t]
\begin{center}
\includegraphics[width=0.46\textwidth]{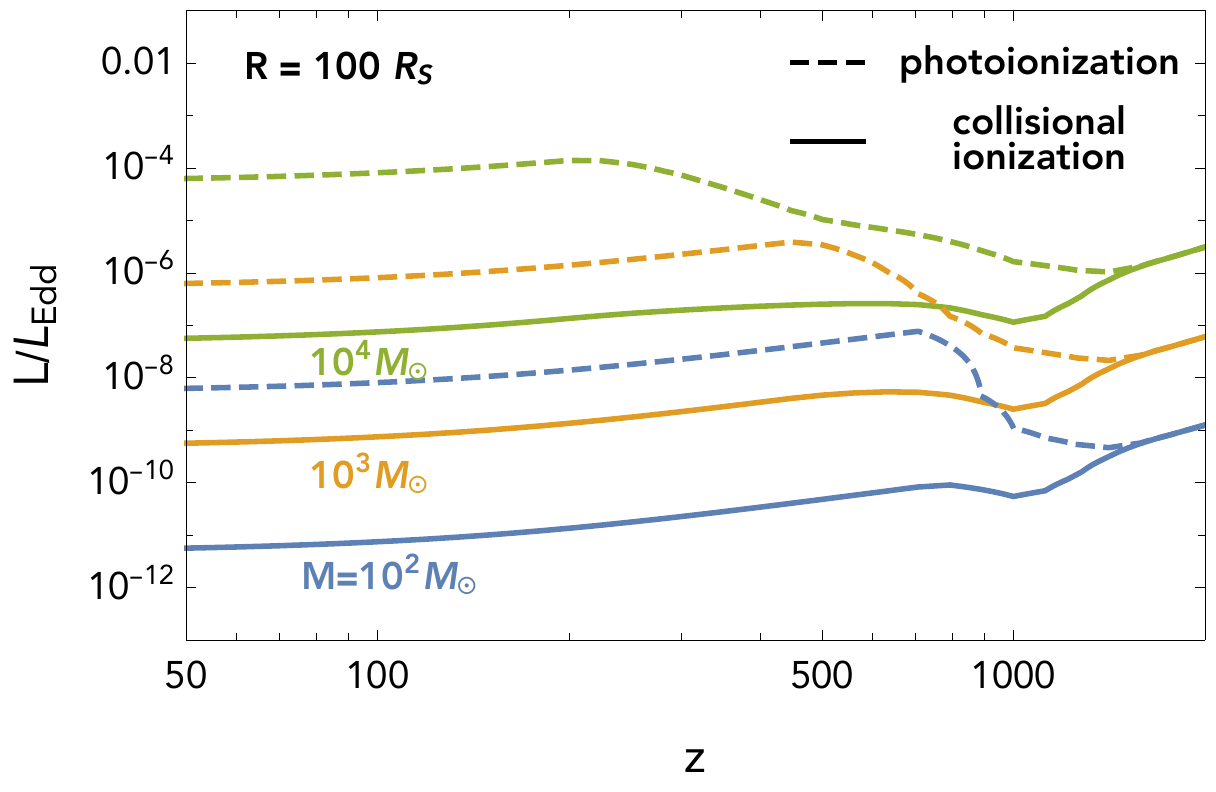} 
\includegraphics[width=0.46\textwidth]{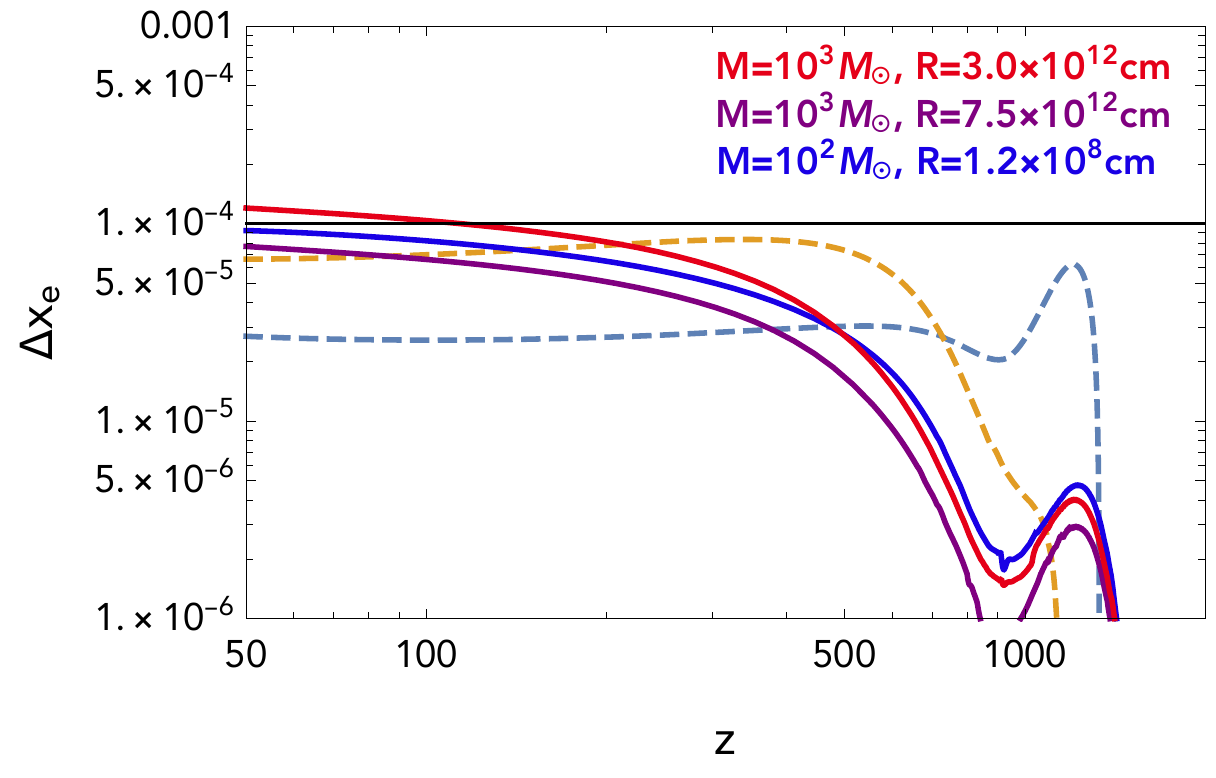} 
\label{fig:Dxe}
\caption{
\textit{Left:} The luminosity of various dMACHOs as a function of redshift, normalized over the Eddington luminosity $L_\mathrm{Edd}=4\pi \GN M m_p/\sigma_\mathrm{T}$. Here, $\sigma_\mathrm{T}$ is the Thomson scattering cross section.  \textit{Right:}  We show the change in the global ionization history $\Delta x_e(z)$ due to the energy injection from hot matter accreted around dMACHOs, which are assumed to make up all the dark matter.  The three sets of masses and radii, $M$ and $R$, are chosen to give $\Delta x_e \approx 10^{-4}$ as $z \to 0$.  For comparison we show the $\Delta x_e$ that results from PBH dark matter with $M = 10^2 \Msun$ and $f_\PBH=1$ (dashed blue curve) and with $M = 10^3 \Msun$ and $f_\PBH = 10^{-2}$ (dashed yellow curve).  All curves in this figures are calculated assuming collisional ionization.
}
\end{center}
\end{figure}

Accretion onto dMACHOs continues into the dark ages, \textit{i.e.} the period of time after recombination at $z \simeq 1100$ and before the ignition of stars at $z \sim O(10)$.  
During this epoch the ionization fraction is small, $x_e \sim 10^{-4} - 10^{-3}$, and the universe is predominantly composed of neutral hydrogen.  
However, if the accreting baryonic matter becomes sufficiently hot, it can also produce radiation with enough energy to ionize the surrounding medium.  
Locally this implies $x_e \to 1$, and when coarse grained on cosmological scales it could imply a shift in the global ionization fraction by as much as $\Delta x_e = O(10^{-4})$ for some benchmark $M$ and $R$ shown later in Fig.~\ref{fig:Dxe}.  
Since ionized gas is less transparent to CMB radiation than neutral hydrogen, an increased $x_e$ affects the visibility function for CMB anisotropies.  
Consequently dMACHO accretion can be constrained by measurements of the CMB power spectra.  

%=========

%=========
Let us first consider how dMACHOs affect the global ionization history, parametrized by $x_e(z)$.  
We previously calculated the radiation power density, $P(z)$ from  \erefs{eq:P_def}{eq:L_ave}, that is emitted from a population of dMACHOs due to their accreted matter.  
Only a fraction of this energy gets deposited into the surrounding thermal bath depending on the efficiency of Compton scattering.  
Let $\dot{\rho}_\mathrm{dep}(t)$ be energy deposition rate (energy per volume per time) at time $t$.  
The energy deposition rate and the energy injection rate are related by~\cite{Ali-Haimoud:2016mbv}
\begin{align}\label{eq:rho_dep_def}
	a^{-7} \frac{d}{dt} \bigl( a^7\dot{\rho}_\mathrm{dep} \bigr) = 0.1 \, n \, \sigma_\mathrm{T} \, \bigl( P - \dot{\rho}_\mathrm{dep} \bigr) \com 
\end{align}
where $n=\rho/m_p=n_H+n_p$ is the {\it overall} density of hydrogen, unionized or ionized, and $\sigma_\mathrm{T} \simeq 6.65 \times 10^{-25} \cm^2$ is the Thomson scattering cross section. It is worth mentioning that we evaluate the energy injection and deposition rates up to $z\sim 2000$. At a higher redshift the CMB dragging force might become important, especially for  heavy dMACHOs. This treatment does not influence our later calculation as the signal of interest is only sensitive to $z\lesssim 1000$.

%=========
To determine the effect of this energy injection on the ionization history, we use the Peebles model~\cite{peebles1968,Poulin:2015pna} plus the additional radiation from dMACHOs to solve for the matter temperature $T_M$ and the ionization fraction $x_e$. 
The coupled equations are written as
\begin{subequations}\label{eq:TM_xe}
\begin{align}
(1+z)H(z)\frac{dT_M}{dz}&=H(z)\left[2\,T_M+\frac{8\pi^2 \,\sigma_\mathrm{T}\,T^4_\mathrm{cmb}}{45\,H(z)\,m_e}\frac{x_e}{1+x_e}(T_M-T_\mathrm{cmb})\right]-\frac{2}{3\,n}\frac{1+2x_e}{3}\dot{\rho}_\mathrm{dep}\,,\\\vspace{5mm}
(1+z)H(z)\frac{dx_e}{dz}&=\frac{1+K_H\Lambda_Hn(1-x_e)}{1+K_H(\Lambda_H+\beta_H)n(1-x_e)}\,\alpha_\mathrm{B}(T_M)\,\left[n\,x^2_e-\left(\frac{m_eT_M}{2\pi}\right)^{3/2}e^{-\frac{E_\mathrm{I}}{T_M}}(1-x_e)\right] \nonumber
\\
&\hspace{5mm} -\frac{1-x_e}{3}\frac{\dot{\rho}_\mathrm{dep}}{E_\mathrm{I}\,n}\,,
\end{align}
\end{subequations}
where $H(z)$ is the Hubble parameter, $\Lambda_H=8.22458 \sec^{-1}$ is the decay rate of the metastable hydrogen 2$S$ state, $K_H=\lambda^3_\mathrm{Ly}/(8\pi H(z))$, $\lambda_\mathrm{Ly}=121.5\, {\rm nm}$ is the wavelength of the Lyman-$\alpha$ photon, and $\alpha_\mathrm{B}$ is the case-B recombination coefficient given below \eref{eq:photo_ionization}. 
The last term in each equation accounts for the additional energy deposition from \eref{eq:rho_dep_def}, without which we should go back to the standard cosmological thermal history.  

%=========
We solve \eref{eq:TM_xe} and present the results in the right panel of \fref{fig:Dxe}.  
For the dMACHO masses and radii that are shown in this figure, the ionization fraction is enhanced by $\Delta x_e \approx 10^{-4}$, which is roughly an $O(1)$ change over the standard calculation without any heating that predicts $x_e \approx 2 \dash 3 \times 10^{-4}$ at $z = 50$. 

%=========
Having understood how dMACHOs affect the global ionization history, the next step is to infer the effect on the CMB anisotropies.  
This could be done by implementing the modified $x_e(z)$ in a Boltzmann code and solving for the CMB power spectra, which has been done in studies of PBH dark matter~\cite{Ali-Haimoud:2016mbv,Blum:2016cjs}.  
However, here we argue that it is not necessary to repeat this Boltzmann analysis, since we can recast existing results for PBHs.  
In particular, we note that the exclusion curves on the $(M, f_\PBH)$ plane derived in \rref{Ali-Haimoud:2016mbv} correspond roughly to ionization histories with $\Delta x_e \approx 10^{-4}$ at $z = 50$.  
This can be seen from Fig.~12 of \rref{Ali-Haimoud:2016mbv}, and we have also reproduced these results in our \fref{fig:Dxe}.  
It is reasonable that the CMB limits would start to be relevant for models that have $\Delta x_e(z=50) \approx 10^{-4}$, since this corresponds to an $O(1)$ change over the prediction without any heating, which gives $x_e \approx 2 \dash 3 \times 10^{-4}$ at $z = 50$.  
All of this goes to say that we will implement the CMB anisotropy constraint by calculating $\Delta x_e(z)$ for different dMACHO masses and radii, and then imposing $\Delta x_e(z=50) < 10^{-4}$.  

\begin{figure}[t]
\begin{center}
\includegraphics[width=0.6\textwidth]{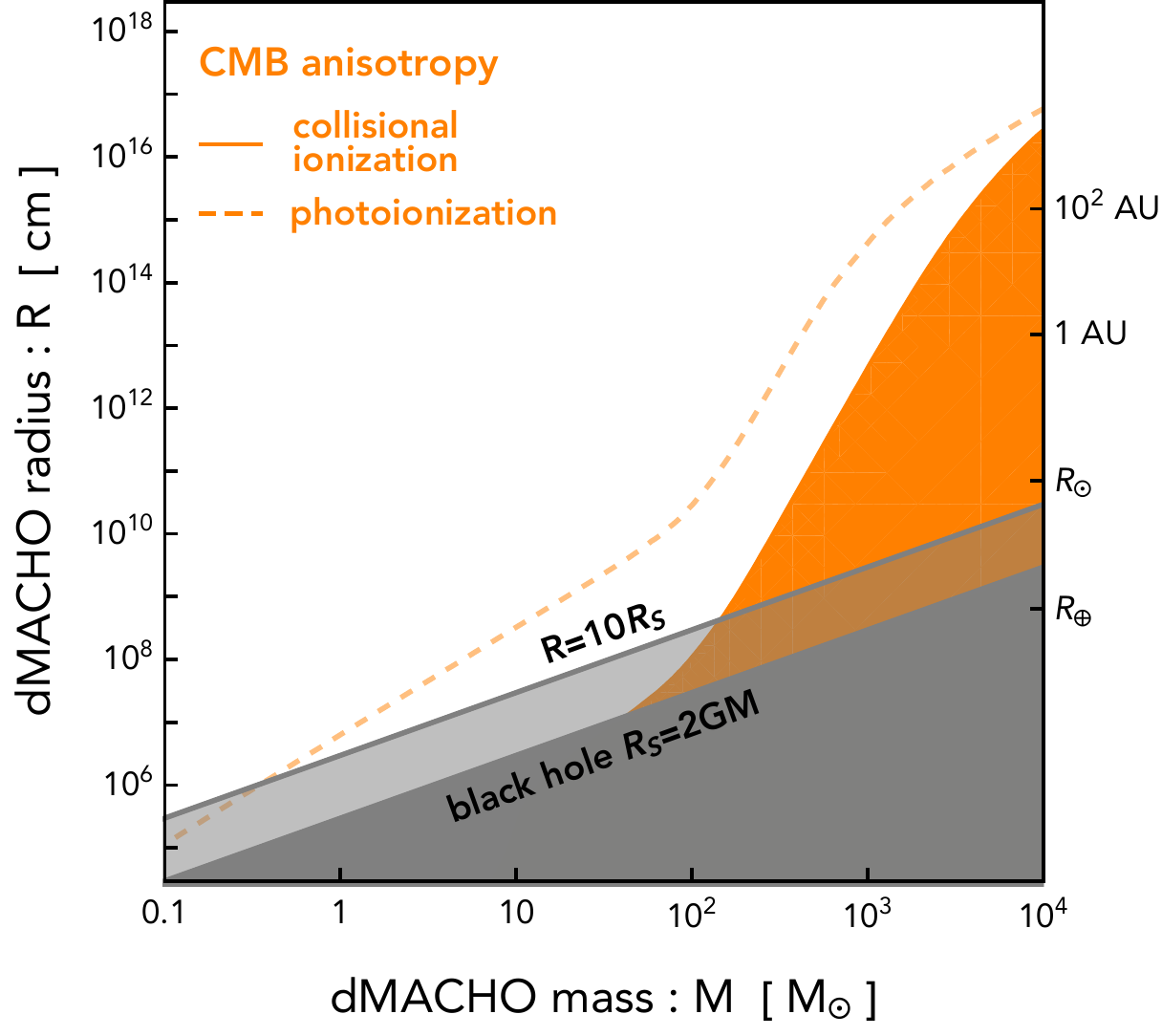} 
\label{fig:CMB_constraint}
\caption{
Constraints on the dMACHO mass-radius parameter space inferred from the change in the cosmic ionization history $\Delta x_e$. We require $\Delta x_e(z=50)<10^{-4}$ such that the CMB anisotropy spectrum is not changed much. Both collisional ionization and photoionization scenarios are considered and shown in the plot.}
\end{center}
\end{figure}

%=========
\fref{fig:CMB_constraint} shows the constraints on the dMACHO parameter space arising from the requirement that there is not too much accretion and ionization so as to disrupt the CMB anisotropies. Specifically for collisional ionization and $M < 10^4\,M_\odot$, the dMACHO radius is constrained to have
\begin{align}
R > (1\times 10^8\,{\rm cm})\times{\rm max}\left[(M/100\,M_\odot)^2,(M/100\,M_\odot)^{9/2}\right] ~,
\end{align}
from CMB anisotropy.
As the dMACHO's mass density approaches that of a black hole, along the upper edge of the gray triangular region, our limit is slightly stronger than the one for PBH dark matter obtained in \rref{Ali-Haimoud:2016mbv}. This difference can be traced back to our use of the hydrostatic approximation (see Fig.~\ref{fig:Hydro_BH_cmp} and the related discussion). However for nearly-critical dMACHOs, corresponding to the light-gray shaded region where $\RS < R \lesssim 10 \RS$, we expect that neither the hydrostatic approximation nor the Bondi approximation provide reliable descriptions of the accretion.

%=========

%==================================
% Glowing dMACHOs in the Milky Way
%==================================
\subsection{Glowing dMACHOs in the Milky Way}\label{sec:current_era}

%=========
In the Milky Way halo today, dMACHOs will accrete the dust and gas that make up our galaxy.  
Here we investigate whether this accretion can lead to such high gas densities and temperatures that the dMACHO develops a glowing halo of baryonic matter.  
In particular we are interested in whether this emission is strong enough to detect with current telescopes.  

%=========
The Milky Way galaxy is vast and varied.  
The ideal environment for efficient accretion onto dMACHOs would involve a low-temperature, high-density, and fully-ionized medium.  
High ionization is preferred, since the accreted matter radiates due to electron-ion scattering.  
Similarly, a higher density means that electron-ion scatterings occur more frequently, and this  increases the emissivity \pref{eq:emissivity}, which grows as $n_e n_p \sim x_e^2 n_H^2$.  
Finally, lower temperature means smaller thermal velocity and more particles are gravitationally-bounded to the dMACHO.  
The Milky Way's interstellar medium (ISM) can be divided into several categories, which are summarized in \tref{tab:ISM}.  
For each ISM environment, this table shows the typical temperature, particle number density, volume filling fraction of the Milky Way, and phase of hydrogen. 
Molecular clouds stand out with their extremely low temperatures and high densities, which makes them good candidates in which to search for glowing dMACHOs.  

%=========
\begin{table}[bth!]
\centering
\label{tab:ISM}
\begin{tabular}{ccccc}
\hline\hline
Medium & Temperature ($\mathrm{K}$) & Density ($\mathrm{cm}^{-3}$) & Vol. Fraction & Hydrogen\\
\hline
Molecular cloud & $\sim 20$ & $\sim 10^3$ & $<1\%$ & molecular\\
\hline
Cold neutral medium & $\sim 100$ & $\sim 20$ & 2-4\% & neutral atomic\\
\hline
Warm neutral medium & $\sim 6000$ & $\sim 0.3$ & $\sim$30\% & neutral atomic\\
\hline
Warm ionized medium & $\sim 8000$ & $\sim 0.3$ & $\sim$15\% & ionized\\
\hline
Hot ionized medium & $\sim 10^6$ & $\sim 10^{-3}$ & $\sim 50\%$ & ionized\\
\hline\hline
\end{tabular}
\caption{A simple summary of the properties of the interstellar media in the Milky Way.  Reproduced from \rref{ism}.
}
\end{table}

\begin{figure}[t] %[tbp]
	\centering
	\includegraphics[width=0.6\linewidth]{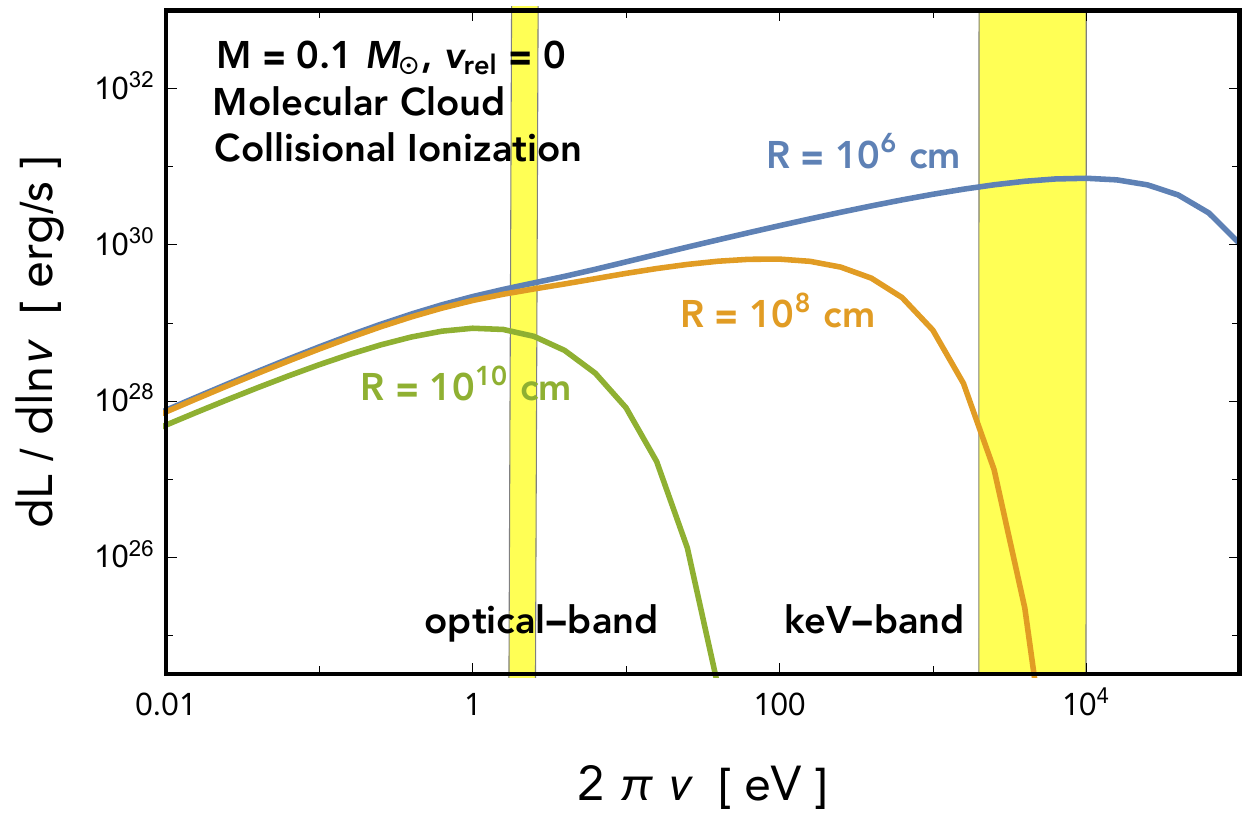} 
	\caption{\label{fig:spectrum} 
	The luminosity spectrum as a function of the radiated photon energy for dMACHOs with $M=0.1\,M_\odot$ and different $R$ in a molecular cloud like Barnard 68. The dMACHO is assumed to be at rest with respect to the molecular cloud.
	}
\end{figure}

%=========
Let us now assess the prospects for seeing a glowing dMACHO assuming that it sits in a cold region of the Milky Way.  
In particular we calculate the luminosity spectrum $L_\nu = dL/d\nu$ using \erefs{eq:L_nu_prime}{eq:L_prime} with the values of $T_\infty$, $\rho_\infty$, and $\bar{x}_e$ from \tref{tab:ISM}.  
The spectra for dMACHOs with mass $M=0.1\,M_\odot$ and several different dMACHO radii are shown in \fref{fig:spectrum}.  

%=========

\begin{figure}[t] %[tbp]
	\centering
	\includegraphics[width=0.46\linewidth]{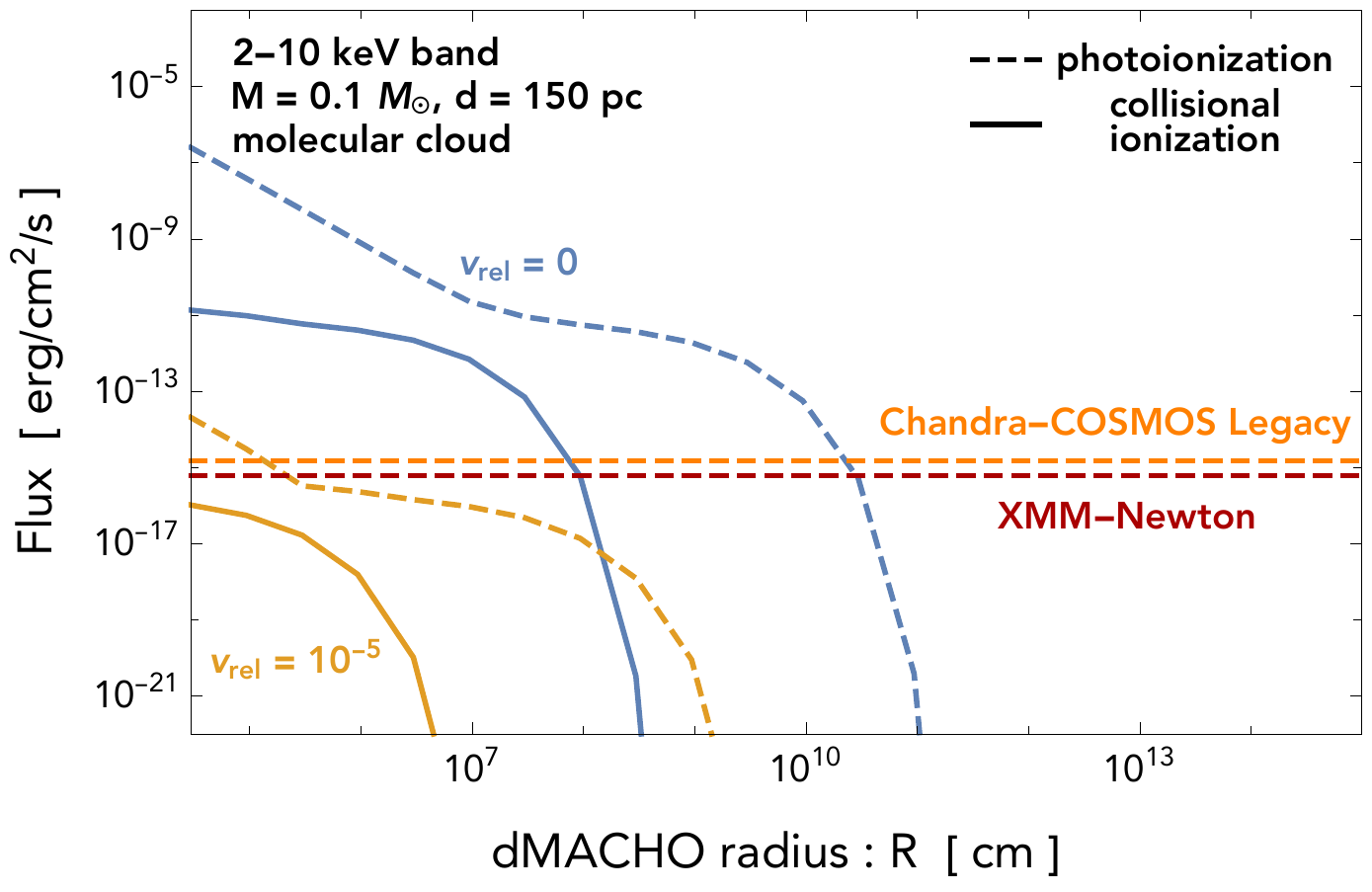} 
	\includegraphics[width=0.46\linewidth]{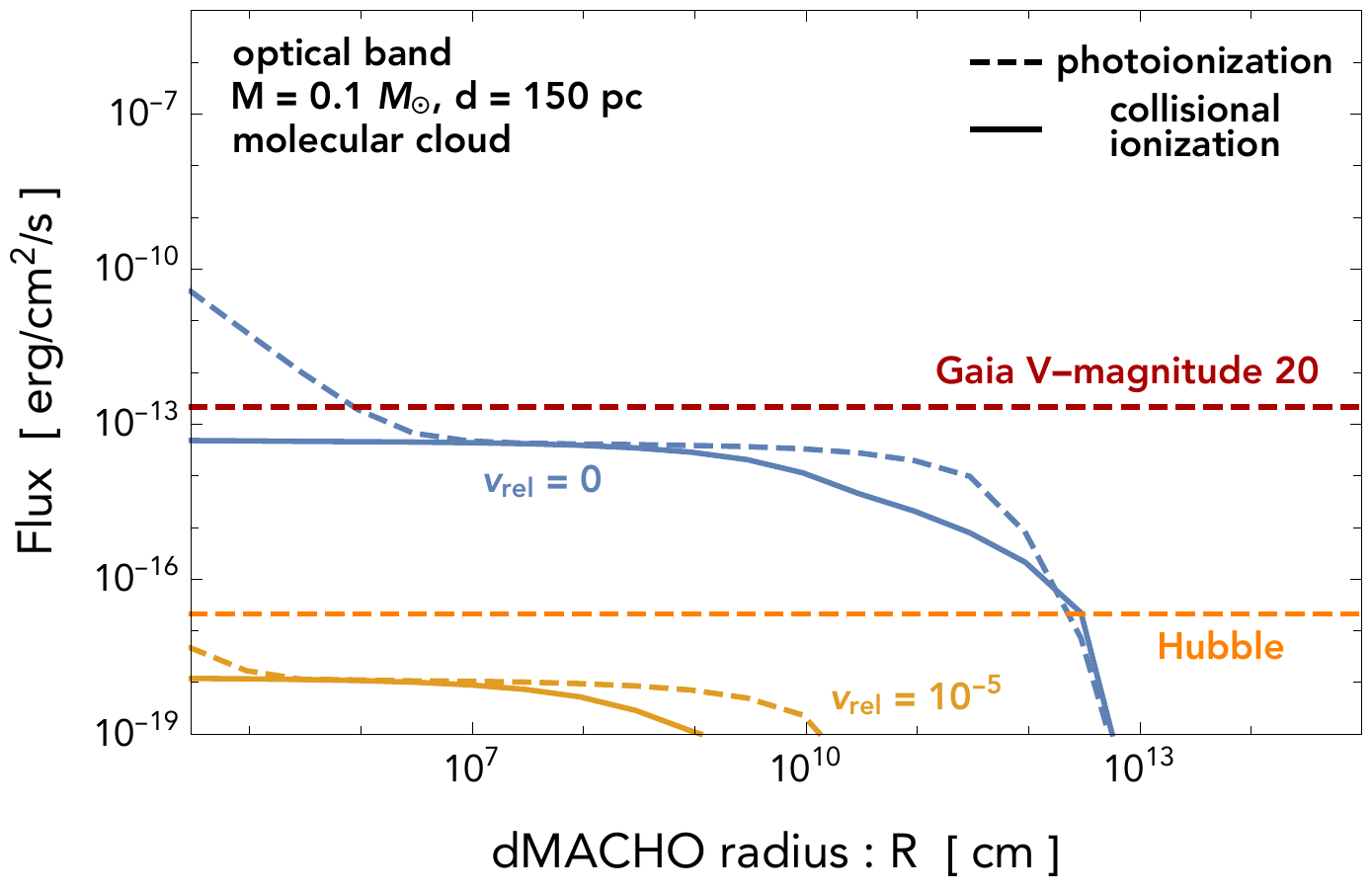} 
	\caption{\label{fig:flux}
	The keV-band (left) and optical band (right) flux from dMACHOs in a molecular cloud. The distance between dMACHO and observer has been chosen to be 150 pc, roughly the distance between Barnard 68 and the Earth. 
The ionization fraction of the molecular $\bar{x}_e$ doesn't influence the result very much, and is explicitly chosen to be $\bar{x}_e=10^{-3}$ in our calculation. 
Also shown in the plot are the flux sensitivities of Chandra-COSMOS Legacy survey~\cite{Civano_2016}, the XMM-Newton telescope~\cite{Brunner:2007kv}, Gaia~\cite{jordi2010gaia} and the Hubble telescope~\cite{Hubble_sensitivity}. The increase of the luminosity at small radius for photoionization is because the temperature at the core of the profile is high enough such that the electrons become relativistic. The fluxes for two different relative velocities, $v_{\rm rel}=0$ and $10^{-5}$, between dMACHO and the molecular cloud are shown here. The probability for dMACHOs in our galaxies to be observed is very small due to the strong suppression from $v_{\rm rel}$, see \eref{eq:P_mc} and \eqref{eq:P_v}.
	}
\end{figure}

%=========
To assess the prospects for detecting this radiation from Earth, we assume that the source is $d = 150\,\pc$ away, which is roughly the distance to the nearby molecular cloud known as Barnard 68~\cite{alves2001internal,hotzel2002kinetic}, and we calculate the flux spectrum as $F_\nu = L_\nu / (4 \pi d^2)$. 
For the sake of illustration we calculate the flux in an optical band from $470 \dash 700 \nm$\footnote{The sensitivity of optical band telescope usually involves a convolution of the raw flux with an acceptance function, {\it e.g.} Gaia's sensitivity is defined with the Johnson-Cousins system~\cite{jordi2010gaia,Bessell_1990}. But for simplicity we ignore this convolution and simply integrate over the wavelength band.} and in a hard X-ray band from $2 \dash 10 \keV$.  
We show the predicted fluxes in \fref{fig:flux} and compare them with the sensitivities of current telescopes.  
We find that a dMACHO with mass $M = 0.1 \Msun$ and small radius $R \lesssim 10^8 \cm$ for collisional ionization ($R \lesssim 10^{10} \cm$ for photoionization) could be detected by X-ray telescopes such as Chandra~\cite{Civano_2016} or XMM-Newton~\cite{Brunner:2007kv} in the $2 \dash 10 \keV$ band when residing in a molecular cloud at rest. 
The former telescope has a sensitivity of $1.5\times 10^{-15}\,{\rm erg/cm^2/s}$ after $\sim$160 ks of effective exposure, and the latter has $9\times 10^{-16}\,{\rm erg/cm^2/s}$ sensitivity after 637 ks of exposure. 
The flux could also be visible to other existing and future X-ray telescopes. For example, the NuSTAR telescope has a sensitivity of $2\times 10^{-15}\,{\rm erg/cm^2/s}$ in the $6 \dash 10 \keV$ band after $10^6$ s of observation \cite{Harrison:2013md}, and the planned eXTP mission has about $2\times 10^{-16}\,{\rm erg/cm^2/s}$ in the $2 \dash 10 \keV$ band after $10^6$ s of exposure~\cite{Zhang:2016ach}. 
Detection prospects are more favorable for photoionization, since the accreted matter can reach a higher temperature and density; see \fref{fig:profiles}.  
We have also checked the optical band, but found that the flux from a $0.1\Msun$ dMACHOs at rest in a molecular cloud is largely below the sensitivity of the Gaia telescope as a point source, and is visible to the Hubble telescope for $R \lesssim 10^{12} \cm$~\cite{Hubble_sensitivity}. 
On the other hand, the flux decreases quickly as the relative velocity between the dMACHO and the molecular cloud increases, as a larger relative velocity leads to a smaller Bondi radius and hence a lower density and temperature at the core of the dMACHO. 
For a dMACHO with $M = 0.1\Msun$, a relative velocity of $v_{\rm rel}\sim 10^{-5}$ is enough to hide its signal from the current telescopes in both the X-ray band and the optical band.

%=========

%=========
Now let us turn our attention to the event rate. 
To estimate the event rate, we need to calculate the probability that a dMACHO encounters a molecular cloud, and also the probability that the velocity of the dMACHO is below the threshold velocity such that its flux is observable to the current telescopes, which implies
\begin{align}
\mathbb{P} = \mathbb{P}_\mathrm{MC} \times \mathbb{P}_{v} ~.
\end{align}
The encounter probability can be estimated as the probability that a dMACHO of mass $M$ resides within a molecular cloud of radius $R_\mathrm{MC}$
\begin{align}
\label{eq:P_mc}
	\Pbb_{\rm MC} \ \approx \ \frac{4\pi}{3} R_\mathrm{MC}^3\,n_\mathrm{dMACHO}\ \simeq \ 33\% \left(\frac{R_\mathrm{MC}}{1\pc}\right)^3\left(\frac{M}{0.1 \Msun}\right)^{-1} \com
\end{align}
where we assume that dMACHOs make up all of the dark matter, and we use the local dark matter density as $\rho_\DM \simeq 0.3 \GeV / \mathrm{cm}^3$.  
The probability for a dMACHO with $M \approx 0.1 \Msun$ to reside in a molecular cloud has been not small.
Heavier dMACHOs could generate a much larger flux, while they are less abundant and thereby reducing the chance that they could show up in a molecular cloud.  
A higher probability is expected with the inclusion of many nearby molecular clouds with a similar high density~\cite{Heyer:2001nz,MCDataBase}.
The probability for the dMACHO and molecular cloud related velocity below the threshold velocity $v^{\rm th}_{\rm rel}$, on the other hand, can be calculated using the three-dimensional Gaussian distribution mentioned in \eref{eq:L_ave} as
\begin{align}
\label{eq:P_v}
\mathbb{P}_v=\frac{1}{(2\pi\langle v^2_\mathrm{L}\rangle/3)^{3/2}} \int d^3\vec{v}\, e^{-\frac{v^2}{2\langle v_\mathrm{L}^2 \rangle/3}}\, \Theta\big( v^{\rm th}_{\rm rel} - |\vec{v} - \vec{v}_\mathrm{MC}|  \big) \,,
\end{align}
where instead of using \eref{eq:vL}, we take $\langle v_\mathrm{L}^2 \rangle^{1/2}=220\,\mbox{km}/\mbox{s}$ as the local dark matter velocity dispersion. Here, $\Theta(x)$ is the Heaviside step function; $\vec{v}_{\rm MC}$ is the molecular cloud velocity in the galaxy frame. 
For $v^{\rm th}_{\rm rel}=10^{-5}$ and $v_\mathrm{MC} \approx 200\,\mbox{km}/\mbox{s}$, we have $\mathbb{P}_v = 1.0\times 10^{-6}$.
Therefore, although dMACHOs have a plausible probability to encounter a molecular cloud close to the Earth, their relative velocities are generically too large to emit enough photon fluxes to be observed by the current telescopes. 

We want to note that our estimation is based on a spherical accretion mechanism, which may not capture the main accretion rate for the case at hand. A non-spherical accretion study for this system may generate a larger luminosity for a generic relative velocity, which we leave for future exploration.

%==================================
% Summary and conclusion
%==================================
\section{Summary and conclusion}\label{sec:conclusion}

%=========
Once we break away from the framework of elementary particle dark matter, a vast landscape of theories becomes accessible.  
The purpose of this work is to provide a phenomenological description of macroscopic, composite dark matter and to survey several strategies for testing these dark matter candidates.  
In general the dMACHO may interact non-gravitationally with visible matter; for example this is the case for quark nuggets or electroweak symmetric dark matter balls~\cite{Ponton:2019hux}.  
However, here we have taken a conservative approach and assumed that the dMACHO interacts only gravitationally, as required by the host of evidence for dark matter's presence in our universe.  
Then we have laid out a roadmap for gravitational tests of dMACHOs.  

%=========
Before we summarize the results of our paper, we want to note that dMACHOs are amenable to a variety of additional probes, which we did not explore in this work. The gravitational lensing of Type Ia supernovae by dMACHOs affects the distribution of perceived luminosities~\cite{Zumalacarregui:2017qqd}.  The window between $1 \dash 100 \Msun$ in the dMACHO $M$-$R$ plane could be probed by this sort of analysis. 
% is also probed by other systems, not considered here, including dwarf galaxies~\cite{Carr:1997cn} and supernovae lensing~\cite{Bernal:2017vvn,Zumalacarregui:2017qqd}.
The gravitational influence of dMACHOs may induce a dynamical friction force on stars in dwarf galaxies leading to constraints similar to earlier work on dark compact objects~\cite{Carr:1997cn}. dMACHOs can also induce distinctive pulsar timing signatures that may be probed by the future Squared Kilometer Arrays~\cite{Dror:2019twh}. The motion of dMACHOs in the Milky Way halo may gravitationally disrupt stellar streams, similar to work on dark matter subhalos~\cite{Bovy:2015mda}.

%=========
Meanwhile, throughout the analysis of our paper we assume a monochromatic distribution of dMACHO masses and radii. However, this assumption is only for the convenience of calculation, and in fact a extended mass function should be expected for dMACHOs. The limits on the parameter space should be varied accordingly, which has been studied in the case of primordial black holes~\cite{Bernal:2017vvn,Kuhnel:2017pwq,Carr:2017jsz}. It would be useful to show how the constraints on dMACHOs change when a extended dMACHO mass and radius function is considered, but this is beyond the purpose of this work.

%=========
The presence of dMACHOs in the Milky Way dark matter halo are expected to induce gravitational lensing of distant stars and galaxies.  
Studies of gravitational lensing, particularly in the context of primordial black holes, tend to assume a point-like lensing mass.  
In this work, we generalize those studies to allow the lensing mass to be distributed in space, and we calculate the corresponding lensing signal as a function of the dMACHO's mass and radius, assuming a uniform density or exponential density profile on the dMACHO's interior.  
(See also \rref{Croon:2020wpr} where the lensing signal is calculated for several density profiles.)  
Our results show that existing PBH lensing constraints from surveys such as Subaru/HSC, OGLE, and EROS/MACHO can be extended into the $M$--$R$ plane out to roughly $R \approx 3\,\RE$ where $\RE$ is the appropriate Einstein radius for a given survey.  
Very large dMACHOs, with $R \gg \RE$, lead to a suppressed lensing and remain unconstrained by these observations. These results are summarized in \fref{fig:mass_radius}. Future lensing surveys using telescopes with exceptional photometric precision can potentially extend the reach by around one order of magnitude in radius with the same sources~\cite{Griest:2011av,Ricker_2014}.

%=========
We have also studied the effects of visible matter accreting on dMACHOs, both in the early universe and today. Accretion is expected to occur in the early universe as a result of the dMACHO's gravitational attractive force.  The accreting matter is heated, which affects the ionization history around the time of recombination, and consequently leaves an imprint of the CMB spectrum and anisotropies.  We study this accretion under the hydrostatic approximation, instead of the Bondi accretion that is often used to study PBH dark matter.  To avoid disrupting the CMB, the dMACHO's mass is bounded from above, and its radius is bounded from below. These measurements provide a robust probe of heavy dMACHOs with $M \gtrsim 100 \Msun$.  Additionally, in today's Universe, dMACHOs in the Milky Way halo may transverse interstellar mediums including molecular clouds. If the dMACHO is almost at rest inside a molecular cloud, the accreted and heated matter can emit X-rays and optical photons, which become detectable for telescopes like Chandra-COSMOS, XMM-Newton, Gaia, and Hubble telescope.

%----------------------------------------------------------------
% Acknowledgements
%----------------------------------------------------------------
\subsubsection*{Acknowledgements}
We would like to thank Yacine Ali-Ha\"imoud and Andrea Isella for discussions of accretion and David Chernoff for suggesting the calculation in \sref{sec:current_era}.  
The work of Y.B. is supported by the U.S. Department of Energy under the contract DE-SC-0017647.  
A.J.L.\ was supported in part by the U.S. Department of Energy under grant DE-SC-0007859.  
A.J.L and Y.B. are grateful to KITP for hospitality during the completion of this work; this research was supported in part by the National Science Foundation under Grant No. NSF PHY-1748958. S.L. is supported in part by Israel Science Foundation under Grant No. 1302/19.

%----------------------------------------------------------------
% References
%----------------------------------------------------------------
\setlength{\bibsep}{6pt}
\bibliographystyle{JHEP}
\bibliography{DarkMACHO}

\end{document}